\documentclass[aps,pra,twocolumn,showpacs,superscriptaddress,preprintnumbers,amsmath,amssymb,floatfix]{revtex4-2}

\usepackage{orcidlink}
\usepackage{xr-hyper}
\usepackage{hyperref} 
\usepackage{amsmath,amssymb} 
\usepackage{subcaption}
\usepackage{mathrsfs}
\usepackage{graphicx}
\usepackage{dcolumn}
\usepackage{bm}
\usepackage{CJK}
\usepackage{color}

\begin{document}

\title{Competing Soft Modes and Tunable Multicriticality in a
	Generalized Two-Mode Quantum Rabi Model}

\author{Xiufeng Cao\,\orcidlink{0000-0003-4500-5772}}
\email{xfcao@xmu.edu.cn}
\affiliation{Department of Physics, Xiamen University,
	Xiamen 361005, China}
\affiliation{Dodd Walls Centre for Photonics and Quantum Technologies and
	Department of Physics, University of Auckland,
	Auckland 1010, New Zealand}
	
\author{Ofri Adiv}
\affiliation{Dodd Walls Centre for Photonics and Quantum Technologies and
	Department of Mathematics, University of Auckland,
	Auckland 1010, New Zealand}

\author{Neil G. R. Broderick}
\affiliation{Dodd Walls Centre for Photonics and Quantum Technologies and
	Department of Physics, University of Auckland,
	Auckland 1010, New Zealand}
 	\date{\today}

			\begin{abstract}
			Multicritical phenomena play a central role in quantum many-body systems, yet their microscopic origin in light--matter platforms remains largely unexplored. Here we investigate a generalized two-mode quantum Rabi model with independently tunable rotating- and counter-rotating-wave couplings. We demonstrate that anisotropy lifts the parent U(1)-symmetric superradiant manifold with a gapless Goldstone-like mode by phase locking the complex superradiant order parameter. This phase-locking mechanism provides the common microscopic origin of the coordinate- and momentum-like soft-mode instabilities, the emergence of symmetry-related two- and four-triple-point multicritical topologies, and the corresponding thermodynamic responses. Within a unified Bogoliubov framework, we further show how the dominant soft mode is redistributed between the two symmetry-related instability channels, thereby determining the multicritical phase structure. The resulting phase boundaries are determined by collective soft-mode softening and coincide exactly with the mean-field instability conditions. We further show that the multicritical topology is directly encoded in the full quantum ground-state energy and its response functions, establishing a unified connection between phase topology, competing collective excitations, and thermodynamic observables. Our results identify competing soft-mode channels as the microscopic origin of tunable multicriticality in strongly coupled light--matter systems.
			
		\end{abstract}

		\maketitle
		
	\section{Introduction}
    
    Quantum phase transitions and critical phenomena constitute one of the central themes of modern many-body physics. In light--matter systems, the Dicke model provides a paradigmatic example, exhibiting a superradiant phase transition from a normal phase to a macroscopically occupied radiation field when the atom--field coupling exceeds a critical value in the thermodynamic limit~\cite{Dicke1954,Hepp1973,Wang1973}. The emergence of superradiant order and collective criticality has stimulated extensive studies across cavity and circuit quantum electrodynamics, trapped ions, and ultracold atomic systems. Subsequent investigations revealed that the Dicke transition is accompanied by characteristic signatures in the excitation spectrum and quantum fluctuations, establishing a close connection between collective soft modes and critical behavior~\cite{Emary2003a,Emary2003b}.
	
    This naturally raises the question of whether the same soft-mode mechanism already emerges in the quantum Rabi model (QRM), the fundamental building block of light--matter interactions. The exact solution of the QRM~\cite{Braak2011} and the subsequent discovery of a genuine quantum phase transition in the large-frequency-ratio limit~\cite{Hwang2015} established the QRM as a minimal platform for exploring quantum criticality beyond conventional many-body settings. This breakthrough triggered extensive investigations of generalized Rabi models, including anisotropic, biased, nonlinear, and open-system extensions~\cite{Liu2017,Hwang2018}. In particular, anisotropic light--matter couplings introduce distinct instability channels associated with different quadratures of the bosonic field and lead to rich critical phenomena beyond the standard superradiant transition~\cite{Liu2017,Xie2025}.
	
	Recent studies have demonstrated that generalized Rabi systems can support multiple competing ground-state instabilities~\cite{Chen2021}, tricritical points and quadruple points~\cite{Ying2021}, as well as multicritical dissipative phase transitions~\cite{Lyu2024}. Experimental quantum simulations have further confirmed the existence of multicritical behavior in both closed and open Rabi models~\cite{Wu2024}, highlighting multicriticality as an emerging frontier of quantum critical phenomena in light--matter systems. More broadly, multicritical points, where several competing phases and instability channels meet simultaneously, play a fundamental role in strongly correlated systems because they provide a natural arena for studying competing orders, emergent symmetries, and collective critical behavior~\cite{Friedemann2018}.
	
	Despite these advances, the microscopic origin and geometric organization of multicritical structures remain incompletely understood. Most existing studies focus on single-mode realizations, where competing instability channels are encoded within a single bosonic degree of freedom~\cite{Chen2021,Ying2021,Lyu2024}. As a result, the relationship between competing soft modes, multicritical topology, and thermodynamic signatures has remained largely unexplored. In particular, it is still unclear whether multiple bosonic modes can fundamentally reorganize the instability landscape and generate qualitatively distinct classes of multicritical topology. Addressing this question requires a framework capable of connecting phase topology, collective excitations, and thermodynamic observables within a unified description.
	
	Multimode light--matter systems provide a natural setting for such investigations. Recent works have shown that additional bosonic degrees of freedom can generate novel collective phenomena, including unconventional superradiant phases, multistability, and complex critical behavior~\cite{RabiTriangle2021,Xu2024,Adiv2024,Zhu2024}. However, the possibility of engineering tunable multicritical topologies through the competition of multiple collective instability channels remains largely unexplored.
	
	In this work, we develop a unified analytical framework for the generalized two-mode quantum Rabi model. We first derive the mean-field steady-state solutions that determine the superradiant order parameters. Building upon these solutions, we formulate Bogoliubov theories for both the normal and superradiant phases, where the latter is obtained through a displaced-frame formulation together with a phase self-consistency condition that eliminates the linear fluctuation terms.
	This framework provides direct analytical access to the collective excitation spectrum, the fluctuation-corrected ground-state energy, and the corresponding thermodynamic response functions within a unified description. It further enables us to establish explicit analytical criteria for the stability of all quantum phases and to determine the complete multicritical phase diagram from the softening of collective modes.
	The resulting theory reveals how spontaneous symmetry breaking, Bogoliubov soft modes, quantum fluctuations, and thermodynamic singularities are interconnected through a common analytical framework, providing a general approach for investigating multicritical quantum phases in strongly coupled light--matter systems.
	
\section{Generalized Two-Mode Anisotropic Quantum Rabi Model and Mean-Field Order Parameters}
\label{sec:MeanField}

\begin{figure}[t]
\centering
\includegraphics[width=0.95\columnwidth]{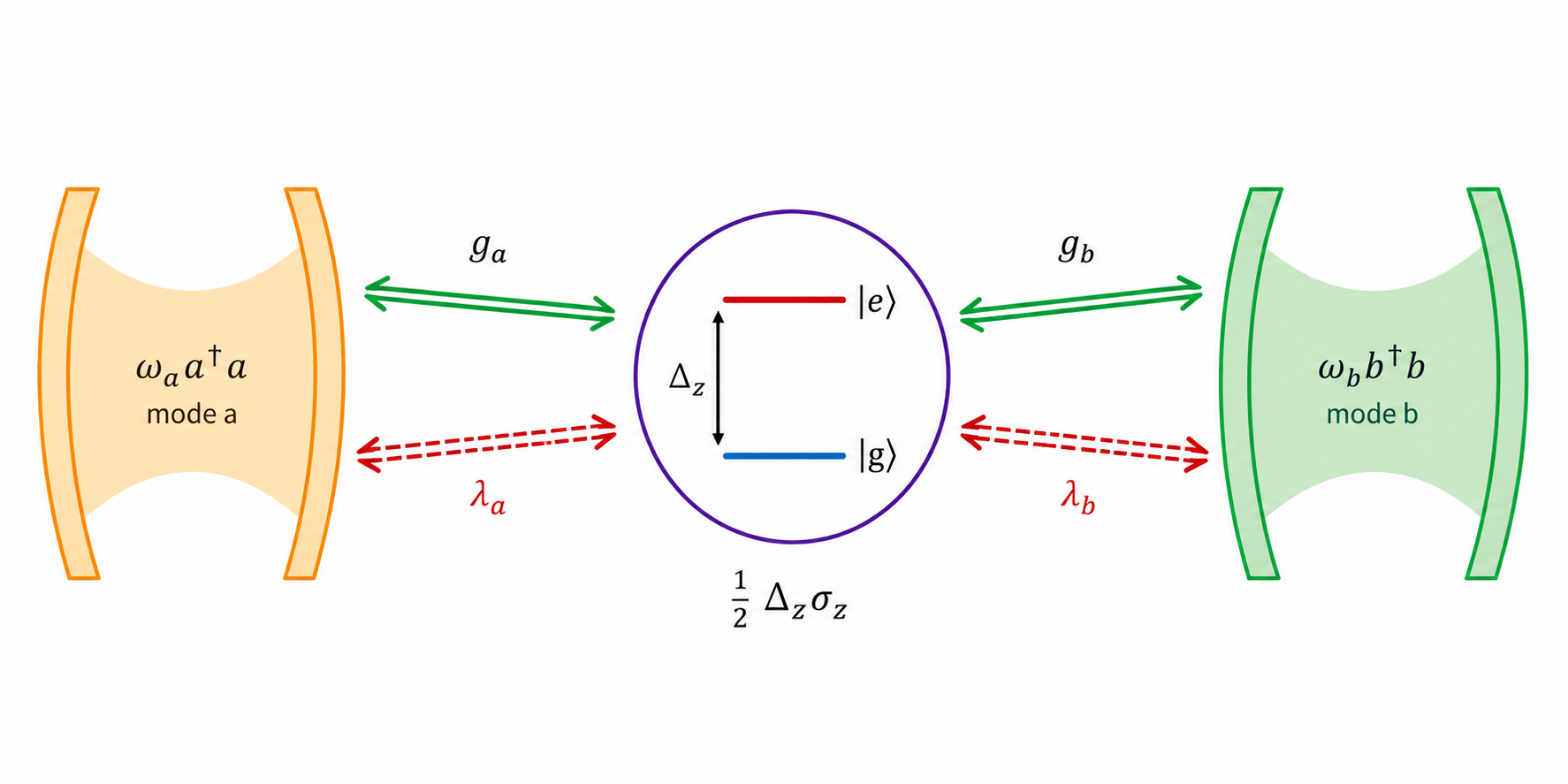}
\caption{Schematic of the generalized two-mode quantum Rabi model.
Two cavity modes with frequencies $\omega_a$ and $\omega_b$ couple to
a two-level system with level splitting $\Delta_z$. Solid (dashed)
arrows denote rotating-wave (counter-rotating-wave) couplings.}
\label{fig:model_schematic}
\end{figure}

We consider a generalized two-mode anisotropic quantum Rabi model in which
a two-level system interacts with two bosonic modes through independently
tunable rotating-wave and counter-rotating-wave couplings. The model is
illustrated schematically in Fig.~\ref{fig:model_schematic}, and its
Hamiltonian is
\begin{equation}
\begin{aligned}
\hat H
={}&
\omega_a\hat a^\dagger\hat a
+
\omega_b\hat b^\dagger\hat b
+
\frac{\Delta_z}{2}\hat\sigma_z
\\
&+
g_a
\left(
\hat a\hat\sigma_+
+
\hat a^\dagger\hat\sigma_-
\right)
+
\lambda_a
\left(
\hat a\hat\sigma_-
+
\hat a^\dagger\hat\sigma_+
\right)
\\
&+
g_b
\left(
\hat b\hat\sigma_+
+
\hat b^\dagger\hat\sigma_-
\right)
+
\lambda_b
\left(
\hat b\hat\sigma_-
+
\hat b^\dagger\hat\sigma_+
\right).
\end{aligned}
\label{eq:Hamiltonian}
\end{equation}
Here, $\omega_{a,b}$ are the bosonic-mode frequencies, $\Delta_z$ is the
two-level splitting, and $g_{a,b}$ and $\lambda_{a,b}$ denote the
rotating-wave and counter-rotating-wave couplings, respectively.

For later convenience, we introduce the coupling combinations
\begin{equation}
A_\pm=g_a\pm\lambda_a,
\qquad
B_\pm=g_b\pm\lambda_b,
\label{eq:ABpm}
\end{equation}
and define the two effective collective coupling strengths
\begin{equation}
\mathcal C_r
=
\frac{A_+^2}{\omega_a}
+
\frac{B_+^2}{\omega_b},
\qquad
\mathcal C_i
=
\frac{A_-^2}{\omega_a}
+
\frac{B_-^2}{\omega_b}.
\label{eq:CrCi}
\end{equation}
The quantities $\mathcal C_r$ and $\mathcal C_i$ characterize the
coordinate-like and momentum-like ordering channels, respectively.
At the mean-field level, the bosonic operators acquire coherent
displacements, $\alpha=\langle a\rangle$, $\beta=\langle b\rangle$.
The self-consistency equations admit two symmetry-distinct superradiant
branches.

For the real-displacement branch, $\mathrm{SR}_r$, the order parameters
are purely real,
\begin{equation}
\alpha_r
=
\pm
\frac{A_+}{2\omega_a}
\sqrt{
1-
\frac{\Delta_z^2}{\mathcal C_r^2}
},
\end{equation}
\begin{equation}
\beta_r
=
\pm
\frac{B_+}{2\omega_b}
\sqrt{
1-
\frac{\Delta_z^2}{\mathcal C_r^2}
}.
\end{equation}
This branch exists for
\begin{equation}
\mathcal C_r\geq\Delta_z,
\label{eq:criticalR}
\end{equation}
with $\mathcal C_r=\Delta_z$ defining the continuous normal-to-$\mathrm{SR}_r$ phase boundary.

For the imaginary-displacement branch, $\mathrm{SR}_i$, the order
parameters are purely imaginary,
\begin{equation}
\alpha_i
=
\pm i
\frac{A_-}{2\omega_a}
\sqrt{
1-
\frac{\Delta_z^2}{\mathcal C_i^2}
},
\end{equation}
\begin{equation}
\beta_i
=
\pm i
\frac{B_-}{2\omega_b}
\sqrt{
1-
\frac{\Delta_z^2}{\mathcal C_i^2}
}.
\end{equation}
This branch exists for
\begin{equation}
\mathcal C_i\geq\Delta_z,
\label{eq:criticalI}
\end{equation}
with $\mathcal C_i=\Delta_z$
defining the continuous normal-to-$\mathrm{SR}_i$ phase boundary.

The complete derivation of the self-consistent order parameters is given
in Appendix~\ref{app:mean-field}. Eqs.~\eqref{eq:criticalR} and
\eqref{eq:criticalI} identify the two independent mean-field instability
channels associated with coordinate-like and momentum-like ordering.
As shown below, the same conditions emerge exactly from the softening of
the corresponding Bogoliubov collective excitations, establishing a
direct correspondence between the mean-field bifurcation and the
collective-mode instability.
	
\section{Effective Low-Energy Hamiltonian in the Normal Phase}
\label{sec:EffectiveNP}

To incorporate the leading quantum fluctuations beyond the mean-field
description, we derive a low-energy effective bosonic Hamiltonian by
perturbatively eliminating the atomic excited state through a
Schrieffer--Wolff (SW) transformation
\cite{Hwang2018,Xie2025,RabiTriangle2021,Zhu2024}.
The SW expansion is controlled in the large-frequency-ratio regime,
$\Delta_z\gg\omega_{a,b}$, with
$g_{a,b}/\Delta_z,\lambda_{a,b}/\Delta_z\ll1$,
where virtual atomic excitations can be eliminated perturbatively while
retaining the leading second-order light--matter contributions that govern
the collective-mode softening. The complete derivation is presented in
Appendix~\ref{app:sw-normalphase}.

Introducing the canonical quadratures
\begin{equation}
x_\mu=\frac{\mu+\mu^\dagger}{\sqrt{2}},
\qquad
p_\mu=\frac{i(\mu^\dagger-\mu)}{\sqrt{2}},
\end{equation}
with $(\mu=a,b)$, the normal-phase effective Hamiltonian takes the form
\begin{equation}
\begin{aligned}
H_{\rm eff}^{N}
=&
-\frac{\Delta_z}{2}
+
\frac{\omega_a}{2}
\left(
x_a^2+p_a^2-1
\right)
+
\frac{\omega_b}{2}
\left(
x_b^2+p_b^2-1
\right)
\\
&-
\frac{1}{2\Delta_z}
\left(
A_+x_a+B_+x_b
\right)^2
-
\frac{1}{2\Delta_z}
\left(
A_-p_a+B_-p_b
\right)^2 .
\end{aligned}
\label{eq:HeffQuadrature}
\end{equation}
Equation~\eqref{eq:HeffQuadrature} explicitly separates the low-energy
fluctuations into coordinate-like and momentum-like collective sectors,
governed by the coupling combinations $(A_+,B_+)$ and $(A_-,B_-)$,
respectively. Their corresponding instability strengths are
$\mathcal{C}_r$ and $\mathcal{C}_i$, introduced in Eq.~\eqref{eq:CrCi}.

The equivalent representation in terms of bosonic creation and annihilation
operators, together with the explicit expressions for the renormalized
coefficients, is given in Appendix~B. The counter-rotating-wave couplings
generate anomalous squeezing and pairing terms in this representation,
leading to particle--hole mixing in the Bogoliubov spectrum. Together with
the number-conserving terms, these quantum fluctuations determine the
competing coordinate-like and momentum-like collective excitations analyzed
below.

\section{Bogoliubov Soft Modes and Ground-State Energy in the Normal Phase}
\label{sec:NormalBdG}

Based on the effective Hamiltonian derived in Sec.~III, we analyze the
collective quantum fluctuations around the normal phase within the
bosonic Bogoliubov--de Gennes (BdG) framework
\cite{RabiTriangle2021,Zhu2024}. Introducing the Nambu spinor
$\Psi=
\begin{pmatrix}
a & b & a^\dagger & b^\dagger
\end{pmatrix}^{T}$,
the quadratic effective Hamiltonian can be cast into the standard BdG
form. The explicit BdG matrix and its subblocks are given in Appendix~\ref{app:sm-normalphase}.

Diagonalization yields two positive collective excitation branches,
whose squared frequencies are
\begin{equation}
\Omega_{N,\pm}^{2}
=
\frac{1}{2}
\left[
T_N
\pm
\sqrt{
T_N^2-4\mathcal{D}_N
}
\right],
\label{eq:SoftMode}
\end{equation}
where
\begin{equation}
T_N
=
{\rm Tr}\!\left(K_p^N K_x^N\right),
\qquad
\mathcal{D}_N
=
\det\!\left(K_p^N K_x^N\right).
\label{eq:TDN}
\end{equation}
The coordinate-sector stiffness matrix is
\begin{equation}
K_x^N
=
\begin{pmatrix}
\displaystyle
\omega_a-\frac{A_+^2}{\Delta_z}
&
\displaystyle
-\frac{A_+B_+}{\Delta_z}
\\[2mm]
\displaystyle
-\frac{A_+B_+}{\Delta_z}
&
\displaystyle
\omega_b-\frac{B_+^2}{\Delta_z}
\end{pmatrix},
\label{eq:KxN}
\end{equation}
while the momentum-sector stiffness matrix is
\begin{equation}
K_p^N
=
\begin{pmatrix}
\displaystyle
\omega_a-\frac{A_-^2}{\Delta_z}
&
\displaystyle
-\frac{A_-B_-}{\Delta_z}
\\[2mm]
\displaystyle
-\frac{A_-B_-}{\Delta_z}
&
\displaystyle
\omega_b-\frac{B_-^2}{\Delta_z}
\end{pmatrix}.
\label{eq:KpN}
\end{equation}
Thus, $K_x^N$ and $K_p^N$ explicitly encode the coordinate-like and
momentum-like collective fluctuation channels, respectively.

The lower branch $\Omega_{N,-}$ controls the stability of the normal
phase. Its softening, $\Omega_{N,-}=0$,
is equivalent to $\det\!\left(K_p^N K_x^N\right)=0$.
Because
\begin{equation}
\det\!\left(K_p^N K_x^N\right)
=
\det K_p^N\,\det K_x^N,
\end{equation}
the instability separates naturally into the coordinate-like and
momentum-like collective channels. The corresponding soft-mode
conditions reduce to $\mathcal{C}_r=\Delta_z$, $\mathcal{C}_i=\Delta_z$,
respectively, in exact agreement with the mean-field critical boundaries
obtained in Sec.~II. This correspondence demonstrates that the onset of
the two superradiant orders is governed by the softening of the
corresponding collective excitation channel.

Geometrically, the two soft-mode conditions define shifted ellipses in
the $(g_a,g_b)$ parameter plane, associated with the coordinate-like and
momentum-like instability channels, respectively. For
$\omega_a=\omega_b=\omega$, they reduce to two circles with the common
radius $\sqrt{\omega\Delta_z}$, centered at $(-\lambda_a,-\lambda_b)$ and $(\lambda_a,\lambda_b),$
respectively. The displacement of the two instability boundaries
therefore provides a direct geometric manifestation of the anisotropy
introduced by the counter-rotating-wave couplings.

Including the zero-point contribution of the two Bogoliubov modes, the
ground-state energy of the normal branch is
\begin{equation}
E_{\rm GS}^{N}
=
\bar E_N
+
\frac{1}{2}
\left(
\Omega_{N,+}
+
\Omega_{N,-}
-
\omega_a
-
\omega_b
\right),
\label{eq:GroundEnergy}
\end{equation}
with $\bar E_N=-\Delta_z/2$.
The second term in Eq.~\eqref{eq:GroundEnergy} is the zero-point
renormalization arising from the collective Bogoliubov fluctuations and
constitutes the leading quantum correction beyond the mean-field energy.

Equations~\eqref{eq:SoftMode} and \eqref{eq:GroundEnergy} thus connect
the collective excitation spectrum, the normal-phase instability, and
the quantum-corrected ground-state energy within the same BdG framework.
They provide the basis for extending the analysis to the superradiant
phases and for identifying the soft-mode competition underlying the
multicritical phase structure.

\section{Unified Phase-Locking Description of Superradiant Ordering}
\label{sec:PhaseLocking}

To describe the ordered phases beyond the normal-state instability, we
introduce the displaced bosonic operators $\hat a=\alpha+\hat c$, $\hat b=\beta+\hat d$,
where the complex amplitudes $\alpha$ and $\beta$ are the superradiant
order parameters, while $\hat c$ and $\hat d$ describe quantum
fluctuations around the displaced mean-field configuration. The detailed
derivation of the displaced-frame Hamiltonian and the associated
self-consistency equations is presented in Appendix~\ref{app:unified-phaselocking}.

In addition to determining the displacement amplitudes, the
self-consistency condition determines the condensate phase $\phi$ through
the phase-locking equation
\begin{equation}
\Delta_z'
=
\frac{\mathcal C_r+\mathcal C_i}{2}
+
\frac{\mathcal C_r-\mathcal C_i}{2}
e^{2i\phi},
\label{eq:PhaseLock}
\end{equation}
where $\mathcal C_r$ and $\mathcal C_i$ are the coordinate-like and
momentum-like collective coupling strengths defined in Sec.~II.
Since $\Delta_z'$ is real, Eq.~\eqref{eq:PhaseLock} requires
\begin{equation}
\left(\mathcal C_r-\mathcal C_i\right)\sin(2\phi)=0,
\label{eq:PhaseCondition}
\end{equation}
which means \begin{equation}
\left(
\frac{g_a\lambda_a}{\omega_a}
+
\frac{g_b\lambda_b}{\omega_b}
\right)
\sin(2\phi)
=
0.
\label{eq:PhaseCondition}
\end{equation}

For $\mathcal C_r\neq\mathcal C_i$, the condensate phase is locked to
four discrete orientations. The solutions $\phi=0,\pi$ correspond to
the real-displacement branch $\mathrm{SR}_r$, whereas
$\phi=\pm\pi/2$ correspond to the imaginary-displacement branch
$\mathrm{SR}_i$. The corresponding order parameters are given explicitly
in Appendix~\ref{app:unified-phaselocking}.

A qualitatively different situation occurs when
$\mathcal C_r=\mathcal C_i$, equivalently,
$g_a\lambda_a/\omega_a+g_b\lambda_b/\omega_b=0$.
For finite counter-rotating-wave couplings, this equality results from
the cancellation of the anisotropy-induced phase-locking term rather
than from restoration of the underlying continuous symmetry. The
$\mathrm{SR}_r$ and $\mathrm{SR}_i$ branches then become degenerate,
and defines the first-order boundary
separating the two superradiant phases.
This accidental degeneracy should be distinguished from the
$U(1)$-symmetric limit, $\lambda_a=\lambda_b=0$. In the latter case,
$\mathcal C_r=\mathcal C_i$ holds identically as a consequence of the
restored continuous symmetry. The condensate phase is then no longer
restricted to discrete orientations, producing a continuously degenerate
superradiant manifold and the associated Goldstone-like collective mode.

Together with the continuous instability conditions
$\mathcal C_r=\Delta_z$, $\mathcal C_i=\Delta_z,$
the degeneracy condition $\mathcal C_r=\mathcal C_i$ completes the
analytical characterization of the phase boundaries. These three
conditions determine, respectively, the normal-to-$\mathrm{SR}_r$
boundary, the normal-to-$\mathrm{SR}_i$ boundary, and the first-order
$\mathrm{SR}_r$-$\mathrm{SR}_i$ boundary. Their intersections generate
the multicritical structure of the phase diagram.

After imposing the self-consistency conditions, all terms linear in the
fluctuation operators vanish, and the displaced Hamiltonian reduces to
\begin{equation}
\begin{aligned}
\hat H_{\rm disp}^{(0)}
={}&
\omega_a|\alpha|^2
+
\omega_b|\beta|^2
+
\omega_a\hat c^\dagger\hat c
+
\omega_b\hat d^\dagger\hat d
-
\frac{\Delta_z}{2}\hat\sigma_z
\\
&+
h_+\hat\sigma_+
+
h_+^*\hat\sigma_-
\\
&+
g_a
\left(
\hat c\hat\sigma_+
+
\hat c^\dagger\hat\sigma_-
\right)
+
\lambda_a
\left(
\hat c\hat\sigma_-
+
\hat c^\dagger\hat\sigma_+
\right)
\\
&+
g_b
\left(
\hat d\hat\sigma_+
+
\hat d^\dagger\hat\sigma_-
\right)
+
\lambda_b
\left(
\hat d\hat\sigma_-
+
\hat d^\dagger\hat\sigma_+
\right).
\end{aligned}
\label{eq:displaced_effective_hamiltonian}
\end{equation}
Eq.~\eqref{eq:displaced_effective_hamiltonian} provides the common
starting point for the fluctuation analysis of the two superradiant
branches, with the displacement amplitudes $\alpha$ and $\beta$
coinciding with the self-consistent mean-field order parameters derived
in Sec.~II. As shown below, the resulting collective excitation spectrum
connects continuously to the normal-phase soft modes and provides a
unified description of the competing instabilities underlying the
multicritical phase structure.

\section{Unified Low-Energy Theory in the Superradiant Phase}

The effective Hamiltonians of the superradiant phases in Eq.~\eqref{eq:displaced_effective_hamiltonian} can be written in the unified form
\begin{equation}
	H_{\rm eff}^{\mu}
	=
	\bar E_{\mu}
	+
	H_{\rm quad}^{\mu},
	\qquad
	\mu={\mathrm{SR}_r},{\mathrm{SR}_i},
	\label{eq:HeffSR}
\end{equation}
where $\bar E_\mu=\omega_a|\alpha|^2 + \omega_b|\beta|^2$ denotes the mean-field energy of the corresponding superradiant branch, while $H_{\rm quad}^{\mu}$ describes the quadratic quantum fluctuations around the self-consistent condensate. Their explicit expressions are summarized in Appendix~\ref{app:lowenergy-superradiantphase}. Since the displacement transformation preserves the quadratic coupling structure of the Hamiltonian, the Bogoliubov analysis in the superradiant phase follows the same diagonalization procedure as that employed in the normal phase.

Although the two superradiant phases possess different condensate order parameters, they differ only in the renormalized fluctuation sector: the coordinate sector in $\mathrm{SR}_{r}$ and the momentum sector in $\mathrm{SR}_{i}$. Both phases therefore share the same low-energy description, differing only through the effective coupling coefficients.

\section{Unified Bogoliubov Theory}
\label{sec:UnifiedBdG}

The low-energy effective Hamiltonians of the normal,
$\mathrm{SR}_r$, and $\mathrm{SR}_i$ phases can be formulated
within a unified bosonic BdG framework, as derived in Appendix~\ref{app:UnifiedBdG}.
In the Nambu spinor $\Psi$,
the effective Hamiltonian of each phase can be written as
\begin{equation}
H_{\rm eff}^{\mu}
=
\bar E_{\mu}
+
\frac{1}{2}
\Psi^\dagger
\mathcal H_{\rm BdG}^{\mu}
\Psi
-
\frac{1}{2}{\rm Tr}(h_\mu),
\label{eq:BdGUnified}
\end{equation}
where $\mu=N,\ \mathrm{SR}_r,\ \mathrm{SR}_i$
labels the normal, real-displacement, and imaginary-displacement
branches, respectively.

The phase dependence is encoded in the coordinate- and
momentum-sector stiffness matrices $K_x^\mu$ and $K_p^\mu$.
Explicitly,
\begin{equation}
K_x^\mu
=
\begin{pmatrix}
\omega_a-C_x^\mu A_+^2
&
-C_x^\mu A_+B_+
\\[2mm]
-C_x^\mu A_+B_+
&
\omega_b-C_x^\mu B_+^2
\end{pmatrix},
\label{eq:KxUnified}
\end{equation}
and
\begin{equation}
K_p^\mu
=
\begin{pmatrix}
\omega_a-C_p^\mu A_-^2
&
-C_p^\mu A_-B_-
\\[2mm]
-C_p^\mu A_-B_-
&
\omega_b-C_p^\mu B_-^2
\end{pmatrix},
\label{eq:KpUnified}
\end{equation}
where the phase-dependent coefficients are
\begin{subequations}
\label{eq:CxpUnified}
\begin{align}
C_x^N
&=
C_p^N
=
\frac{1}{\Delta_z},
\label{eq:CxpN}
\\
C_x^{\mathrm{SR}_r}
&=
\frac{\cos^2\theta_r}{\Delta_r},
\qquad
C_p^{\mathrm{SR}_r}
=
\frac{1}{\Delta_r},
\label{eq:CxpSRr}
\\
C_x^{\mathrm{SR}_i}
&=
\frac{1}{\Delta_i},
\qquad
C_p^{\mathrm{SR}_i}
=
\frac{\cos^2\theta_i}{\Delta_i}.
\label{eq:CxpSRi}
\end{align}
\end{subequations}

Diagonalization of Eq.~\eqref{eq:BdGUnified} yields two positive
collective excitation branches,
\begin{equation}
\Omega_{\mu,\pm}^{2}
=
\frac{1}{2}
\left[
T_\mu
\pm
\sqrt{
T_\mu^{2}
-
4D_\mu
}
\right],
\label{eq:SoftUnified}
\end{equation}
where
\begin{equation}
T_\mu
=
{\rm Tr}\!\left(K_p^\mu K_x^\mu\right),
\qquad
D_\mu
=
\det\!\left(K_p^\mu K_x^\mu\right).
\end{equation}
The lower branch $\Omega_{\mu,-}$ controls the stability of each
phase, and its softening, $\Omega_{\mu,-}=0,$
signals the corresponding collective-mode instability.

Including the zero-point contribution of the Bogoliubov modes,
the ground-state energy of each branch is
\begin{equation}
E_{\rm GS}^{\mu}
=
\bar E_{\mu}
+
\frac{1}{2}
\left(
\Omega_{\mu,+}
+
\Omega_{\mu,-}
-
\omega_a
-
\omega_b
\right).
\label{eq:BranchGround}
\end{equation}
The physical ground-state energy is then determined by the global
minimum among the three competing branches,
\begin{equation}
E_{\rm GS}
=
\min
\left\{
E_{\rm GS}^{N},
E_{\rm GS}^{\mathrm{SR}_r},
E_{\rm GS}^{\mathrm{SR}_i}
\right\}.
\label{eq:GlobalGround}
\end{equation}
This unified formulation places the normal and the two
superradiant phases on the same footing: their stability is
determined by the corresponding collective excitation spectrum,
while the physical phase is selected by the global minimum of
the BdG-corrected ground-state energy.

\section{Soft-Mode Instabilities}

The soft-mode boundaries are determined by the vanishing of the lower
Bogoliubov branch, $\Omega_{\mu,-}=0$, which is equivalent to
$D_\mu=0$, as derived in Appendix~\ref{app:SoftModeConditions}. In terms of the instability strengths
$\mathcal{C}_r$ and $\mathcal{C}_i$ defined in Eq.~\eqref{eq:CrCi}, the
soft-mode condition for all three phases can be written in the unified form
\begin{equation}
\left(1-C_x^\mu\mathcal{C}_r\right)
\left(1-C_p^\mu\mathcal{C}_i\right)
=0.
\label{eq:UnifiedSoft}
\end{equation}
The phase-dependent coefficients $C_x^\mu$ and $C_p^\mu$ are given in Eq.~\eqref{eq:CxpUnified}.

Equation~\eqref{eq:UnifiedSoft} shows that the stability of all three
phases is governed by the same two competing collective channels.
In the normal phase, the coordinate-like and momentum-like channels
enter on an equal footing, whereas the $\mathrm{SR}_r$ and
$\mathrm{SR}_i$ phases selectively renormalize the coordinate and
momentum sectors, respectively. The redistribution of these two
soft-mode channels determines the multicritical structure of the
phase diagram.

\section{Result and Discussion}

\begin{figure*}[t]
	\centering
	\includegraphics{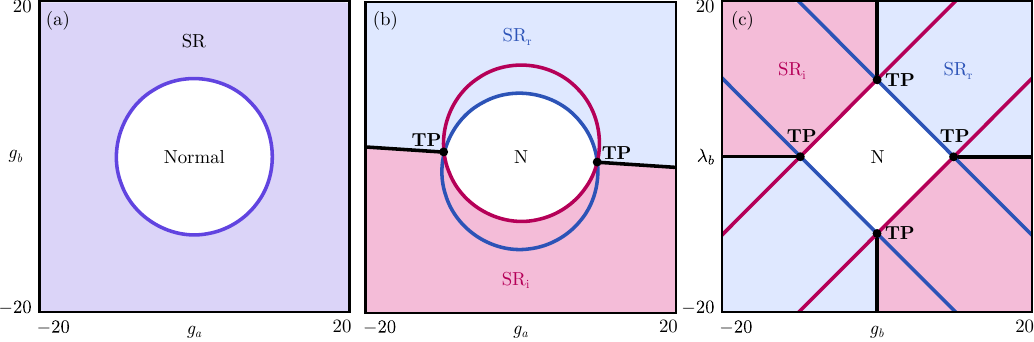}
	\caption{Emergence of tunable multicritical topologies in the generalized two-mode quantum Rabi model.
		(a) In the $U(1)$-symmetric limit ($\lambda_a=\lambda_b=0$), the $\mathrm{SR}_{r}$ and $\mathrm{SR}_{i}$ phases merge into a continuously degenerate superradiant manifold bounded by a circular soft-mode instability line, supporting a Goldstone-like collective mode.
		(b) Finite anisotropy ($\lambda_a=0.1$, $\lambda_b=1.8$) lifts the Goldstone degeneracy and generates a two-triple-point topology in the $(g_a,g_b)$ plane.
		(c) Under a different parameter section ($g_a=0.1$, $\lambda_a=1.8$), the competing superradiant orders reorganize into a diamond-shaped four-triple-point topology in the $(g_b,\lambda_b)$ plane.
		Blue and magenta curves denote continuous soft-mode boundaries, black solid curves indicate first-order $\mathrm{SR}_{r}$/$\mathrm{SR}_{i}$ transitions, and black dots mark triple points where the normal, $\mathrm{SR}_{r}$, and $\mathrm{SR}_{i}$ phases meet.
	}
	\label{fig:PhaseDiagrams}
\end{figure*}

Figure~\ref{fig:PhaseDiagrams} illustrates how the phase diagram of the generalized two-mode quantum Rabi model evolves from a symmetry-protected Goldstone phase into distinct multicritical topologies as the anisotropy parameters are tuned. In the U(1)-symmetric limit shown in Fig.~\ref{fig:PhaseDiagrams}(a), where $\lambda_a=\lambda_b=0$, the two quadrature channels become exactly degenerate and the distinction between the $\mathrm{SR}_{r}$ and $\mathrm{SR}_{i}$ phases disappears. The normal-to-superradiant transition is determined by the soft-mode condition $\frac{g_a^2}{\omega_a}+\frac{g_b^2}{\omega_b} = \Delta_z$. For $\omega_a=\omega_b=1$ and $\Delta_z=100$, the critical boundary reduces to the circle $g_a^2+g_b^2=100$. Inside the circle the system remains in the normal phase, whereas outside the circle a continuously degenerate superradiant manifold emerges. The spontaneous breaking of the continuous U(1) symmetry generates a gapless Goldstone-like collective mode. This symmetry-protected Goldstone phase serves as the parent structure of the multicritical topologies that appear once anisotropy is introduced.

Figure~\ref{fig:PhaseDiagrams}(b) shows the phase diagram in the $(g_a,g_b)$ plane for finite anisotropy, with $\lambda_a=0.1$ and $\lambda_b=1.8$. The U(1) degeneracy is lifted, splitting the superradiant manifold into the competing SR$_r$ and SR$_i$ phases. In Fig.~2(b), the continuous soft-mode boundaries intersect the first-order SR$_r$--SR$_i$ boundary at two symmetry-related triple points, where the normal and both superradiant phases coexist. This two-triple-point topology reflects the competition between coordinate- and momentum-like ordering channels. A distinct structure appears in Fig.~2(c), where $g_a=0.1$ and $\lambda_a=1.8$ are fixed while $(g_b,\lambda_b)$ are varied. The nearly linear soft-mode boundaries form a diamond-shaped normal region with four triple points at its vertices, while the $SR_r$ and $SR_i$ phases occupy alternating sectors. 

Taken together, Figs.~2(a)–(c) demonstrate that the multicritical topology is continuously tunable through the coupling anisotropies. The evolution from the U(1)-symmetric circular boundary to the two- and four-triple-point structures reflects the continuous reorganization of the coordinate- and momentum-like soft-mode channels. The microscopic origin of this tunable multicriticality is elucidated in the following sections through a unified Bogoliubov analysis.

\subsection{Dynamical Origin of Tunable Multicritical Topologies}

\begin{figure*}[!t]
    \centering

    \begin{minipage}{\textwidth}
        \centering
        \includegraphics[width=\textwidth]{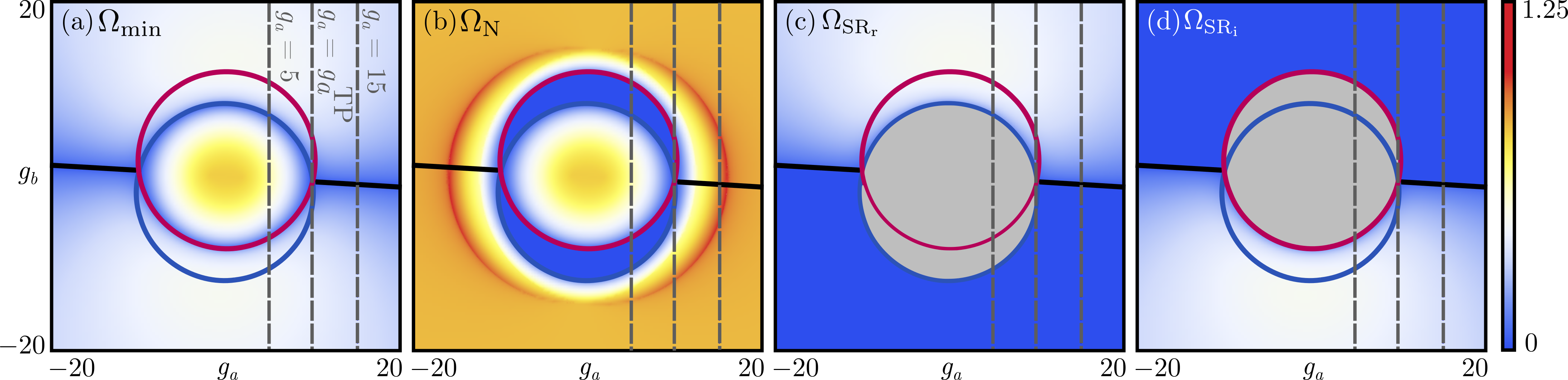}

        \captionof{figure}{
       Competing soft-mode structure underlying the two-triple-point topology of the generalized two-mode quantum Rabi model.
		(a) Global physical soft mode selected by the ground-state phase.
		(b) Normal-phase soft mode.
		(c) $\mathrm{SR}_{r}$ soft mode.
		(d) $\mathrm{SR}_{i}$ soft mode.
		Blue and magenta curves denote continuous soft-mode instability boundaries,
		and the black solid line denotes the first-order boundary separating the $\mathrm{SR}_{r}$ and $\mathrm{SR}_{i}$ phases.
		The two triple points arise from the competition between coordinate-like and momentum-like instability channels.
		The redistribution of the dominant soft mode among the normal, $\mathrm{SR}_{r}$, and $\mathrm{SR}_{i}$ phases establishes the dynamical mechanism responsible for the multicritical phase structure.
        }
        \label{fig:TwoTPsoftmode_maps}
    \end{minipage}

    \vspace{0.8em}

    \begin{minipage}{0.94\textwidth}
        \centering
        \includegraphics[width=\textwidth]{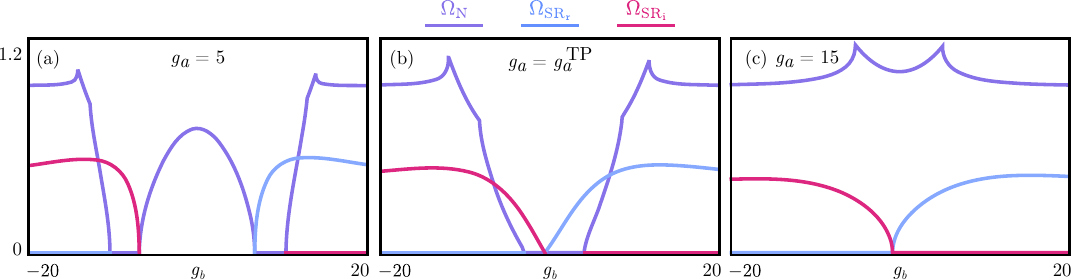}

        \captionof{figure}{
        Representative one-dimensional cuts of the soft-mode spectrum across the two-triple-point topology.
		(a) Normal-side cut ($g_a=5$), where the soft mode softens only at the continuous normal-to-superradiant transition.
		(b) Triple-point cut ($g_a=g_a^{\mathrm{TP}}\approx9.821$), where the competing instability channels simultaneously approach criticality.
		(c) Superradiant-side cut ($g_a=15$), illustrating the evolution of the collective excitation spectrum deep inside the ordered phase.
		The comparison reveals how the dominant soft mode is redistributed among the normal, $\mathrm{SR}_{r}$, and $\mathrm{SR}_{i}$ sectors as the system passes through the multicritical region.
        }
        \label{fig:TwoTPSoftModeCuts}
    \end{minipage}

\end{figure*}

Figure~\ref{fig:TwoTPsoftmode_maps} reveals the dynamical origin of the multicritical topologies identified in Fig.~\ref{fig:PhaseDiagrams}. The central result is that the phase diagram is governed not by a single superradiant instability, but by the competition between two Bogoliubov soft-mode channels associated with the coordinate- and momentum-like collective fluctuations. This competition provides a unified dynamical framework for understanding the phase boundaries and the emergence of the multicritical topology.

Figure~\ref{fig:TwoTPsoftmode_maps}(a) presents the global minimum soft mode, $\Omega_{\min}$,
obtained by selecting the lowest excitation among the normal, SR$_r$, and SR$_i$ phases throughout the $(g_a,g_b)$ parameter space. The continuous phase boundaries are determined by the softening of one of the two Bogoliubov modes, whereas the first-order SR$_r$-SR$_i$ boundary separates the two superradiant phases. The two triple points are located at the intersections of the continuous and first-order boundaries. Three representative cuts, $g_a=5$, $g_a=g_a^{\rm TP}$, and $g_a=15$, are chosen to illustrate how the excitation spectrum evolves across different regions of the phase diagram.

The phase-resolved spectra in Figs.~\ref{fig:TwoTPsoftmode_maps}(b)–~\ref{fig:TwoTPsoftmode_maps}(d) further illustrate how the dominant soft mode evolves across the normal and superradiant phases. In the normal phase [Fig.~\ref{fig:TwoTPsoftmode_maps}(b)], the soft mode vanishes at the normal-to-superradiant boundary, showing that the onset of superradiance is driven by a normal-phase instability. In contrast, Figs.~\ref{fig:TwoTPsoftmode_maps}(c) and ~\ref{fig:TwoTPsoftmode_maps}(d) show that the SR$_r$ and SR$_i$ soft modes dominate complementary regions of the ordered phase. Across the first-order boundary, the dominant soft mode switches between the SR$_r$ and SR$_i$ branches, reflecting the exchange of stability between the two superradiant phases. This redistribution of the dominant soft mode provides the dynamical mechanism underlying the multicritical topology.

Figure 4 presents representative one-dimensional cuts of the soft-mode spectra at fixed $g_a$. These cuts trace the evolution of the competing Bogoliubov soft modes from an ordinary transition $g_a=5$ in Fig.~4(a), through the multicritical point $g_a=g_a^{\rm TP}\simeq 9.821$ Fig.~4(b), and into the superradiant regime $g_a=15$  Fig.~4(c).
Away from the multicritical region in Fig. 4(a), the $\mathrm{SR}_{r}$
and $\mathrm{SR}_{i}$ instability channels soften at well-separated values of $g_b$. The normal phase excitation therefore approaches zero only at isolated critical points, indicating that the transition is governed by a single dominant instability channel. At the triple point, $g_a=g_a^{\rm TP}\simeq 9.821$ in Fig. 4(b), the two instability channels approach criticality simultaneously. The coordinate- and momentum-sector collective modes become nearly degenerate, producing direct competition between the two soft modes. This simultaneous softening constitutes the dynamical signature of the multicritical point and corresponds to the intersection structure in Fig. 3(a). Deep inside the superradiant regime in Fig.~\ref{fig:TwoTPSoftModeCuts}(c), the competition is resolved through branch selection. The $\mathrm{SR}_{i}$ mode dominates for $g_b<-\frac{g_a\lambda_a\omega_b}{\omega_a\lambda_b} \approx-0.83$, whereas the $\mathrm{SR}_{r}$ mode dominates for $g_b>-0.83$. The first-order boundary separates the two ordered sectors, reflecting the stabilization of one collective channel at the expense of the other.

Together, Figs.~3 and 4 provide complementary views of the same physical mechanism. 
The two-dimensional maps in Fig.~3 identify the spatial organization of the soft-mode instabilities across the phase diagram, whereas the cuts in Fig.~4 resolve how the competing excitation branches evolve when the system passes through ordinary, multicritical, and superradiant regimes. Their combined analysis establishes the competition between Bogoliubov soft modes as the microscopic origin of the tunable multicritical topology.

\begin{figure*}[!t]
    \centering

    \begin{minipage}{\textwidth}
        \centering
        \includegraphics[width=\textwidth]{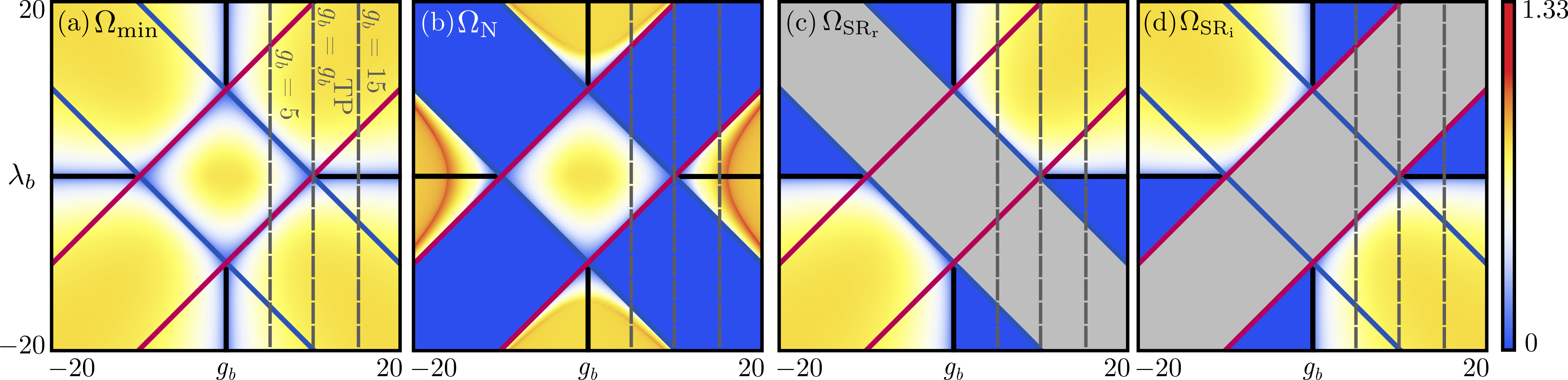}
        \captionof{figure}{Soft-mode structure in the $(g_b,\lambda_b)$ plane.
		(a) Global minimum soft mode $\Omega_{-}$.
		(b) Normal-phase soft mode $\Omega_{N,-}$.
		(c) Superradiant-r soft mode $\Omega_{{\rm \mathrm{SR}_{r}},-}$.
		(d) Superradiant-i soft mode $\Omega_{{\rm \mathrm{SR}_{i}},-}$.
        }
        \label{fig:FourTPsoftmode_maps}
    \end{minipage}

    \vspace{0.8em}

    \begin{minipage}{0.94\textwidth}
        \centering
        \includegraphics[width=\textwidth]{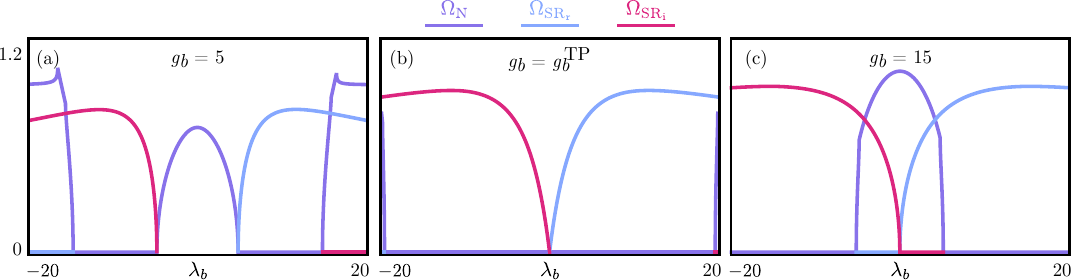}

        \captionof{figure}{
        Representative one-dimensional cuts of the soft-mode spectrum
        across the four-triple-point topology.
        (a) Normal-side cut ($g_b=5$), where the soft mode softens only at
        the continuous normal-to-superradiant transition.
        (b) Triple-point cut ($g_b=g_b^{\mathrm{TP}}$), where the competing
        instability channels simultaneously approach criticality.
        (c) Superradiant-side cut ($g_b=15$), illustrating the evolution
        of the collective excitation spectrum deep inside the ordered phase.
        }
        \label{fig:FourTPSoftModeCuts}
    \end{minipage}

\end{figure*}

To elucidate the microscopic origin of the four-triple-point topology, Figs.~\ref{fig:FourTPsoftmode_maps} and ~\ref{fig:FourTPSoftModeCuts} provide complementary views of the competing collective excitations in the $(g_b,\lambda_b)$ plane.
Figure~\ref{fig:FourTPsoftmode_maps}(a) displays the global minimum excitation $\Omega_-$. The soft-mode boundaries form a diamond-shaped network enclosing the normal phase, in agreement with the phase diagram shown in Fig.~2(c). Four symmetry-related triple points appear at the vertices of the diamond and are connected pairwise by the first-order boundaries determined by $g_a\lambda_a/\omega_a+g_b\lambda_b/\omega_b=0$.
Compared with the circular topology of the two-triple-point case, the competing collective excitations are redistributed into four symmetry-related sectors, giving rise to a qualitatively different multicritical geometry.

The decomposition into individual excitation channels is shown in Figs.~\ref{fig:FourTPsoftmode_maps}(b)-\ref{fig:FourTPsoftmode_maps}(d). Figure~\ref{fig:FourTPsoftmode_maps}(b) presents the normal-phase excitation, whose gap continuously decreases toward the diamond boundaries and vanishes along the four critical lines, signaling the onset of the normal-to-superradiant instability. Figures~\ref{fig:FourTPsoftmode_maps}(c) and ~\ref{fig:FourTPsoftmode_maps}(d) display the coordinate-like and momentum-like collective excitations inside the ordered phases, respectively. Instead of occupying opposite sides of the phase diagram as in the two-triple-point topology, the two excitation channels alternate around the normal phase, forming four disconnected sectors separated by the first-order boundaries. This fourfold redistribution of competing collective excitations directly reflects the multicritical organization of the phase diagram.

Figure~\ref{fig:FourTPSoftModeCuts} further resolves the evolution of the low-energy spectrum along representative one-dimensional cuts. Away from the multicritical region Fig.~~\ref{fig:FourTPSoftModeCuts}(a), the normal excitation softens only at isolated normal-to-superradiant transition points, indicating that each transition is governed by a single instability channel. At the representative triple-point cut in Fig.~\ref{fig:FourTPSoftModeCuts}(b), the coordinate-like and momentum-like collective excitations simultaneously approach criticality, becoming nearly degenerate over the same parameter region. This simultaneous softening constitutes the dynamical signature of the multicritical point. Deep inside the superradiant regime in Fig.~\ref{fig:FourTPSoftModeCuts}(c), the multicritical competition is resolved through branch selection across the first-order boundary, where one collective excitation is stabilized while the competing channel becomes gapped.

Together, Figs.~\ref{fig:FourTPsoftmode_maps} and
\ref{fig:FourTPSoftModeCuts} demonstrate that the four-triple-point
topology originates from the redistribution and competition of the
coordinate-like and momentum-like collective excitations across four
symmetry-related sectors, with the resulting diamond-shaped soft-mode
network providing its direct dynamical manifestation.

The comparison between Figs.\ref{fig:TwoTPsoftmode_maps} and \ref{fig:FourTPsoftmode_maps} shows that the
multicritical topology is not an independent geometric
feature of the phase diagram, but is generated by the
parameter-space organization of the competing
coordinate-like and momentum-like collective excitations.
The corresponding continuous instabilities are determined by
$\mathcal{C}_r=\Delta_z$ and $\mathcal{C}_i=\Delta_z$, while the
first-order boundary separating the two ordered branches satisfies
$g_a\lambda_a/\omega_a+g_b\lambda_b/\omega_b=0$, or equivalently,
$\mathcal{C}_r=\mathcal{C}_i$. The triple points therefore occur where
$\mathcal{C}_r=\mathcal{C}_i=\Delta_z$. In the two-triple-point topology,
these conditions yield only one symmetry-related pair of multicritical
points, so that the competition between the two collective excitation
channels remains localized near these points. In the
four-triple-point topology, the same two channels are
redistributed among four symmetry-equivalent sectors,
producing four intersections and the associated
diamond-shaped network of soft-mode boundaries. The
number and arrangement of triple points are thus controlled
by the geometry of the intersections between the continuous
soft-mode boundaries and the first-order degeneracy
condition in the chosen parameter plane.

\subsection{Ground-State Energy Signatures of Multicriticality}

Since the multicritical topology originates from the competition between
coordinate-like and momentum-like collective excitations, it is natural to
ask how this competition manifests itself in thermodynamic observables.
Figures~7-10 address this question through the ground-state energy and its
derivatives, establishing the thermodynamic counterpart of the soft-mode
topology.

\begin{figure}
	\centering
	\includegraphics{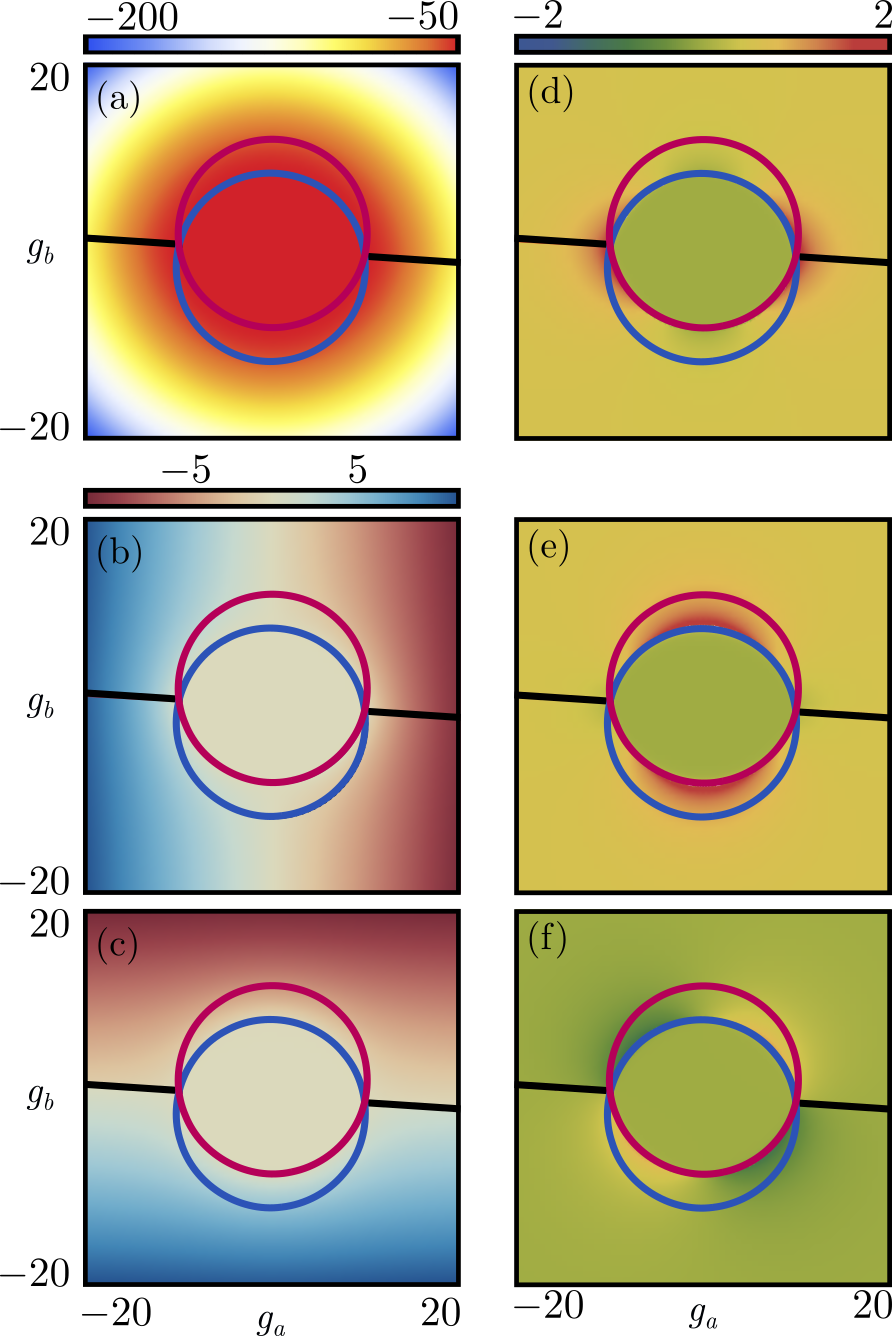}
	\caption{
		Ground-state energy and thermodynamic response functions in the $(g_a,g_b)$ plane.
		(a) Ground-state energy $E_{\rm GS}$.
		(b) First derivative $\partial E_{\rm GS}/\partial g_a$.
		(c) First derivative $\partial E_{\rm GS}/\partial g_b$.
		(d) Response function $\chi_{g_a g_a}$.
		(e) Response function $\chi_{g_b g_b}$.
		(f) Mixed response function $\chi_{g_a g_b}$.
	}
	\label{fig:energy_response_ga_gb}
\end{figure}

\begin{figure}
	\centering
	\includegraphics{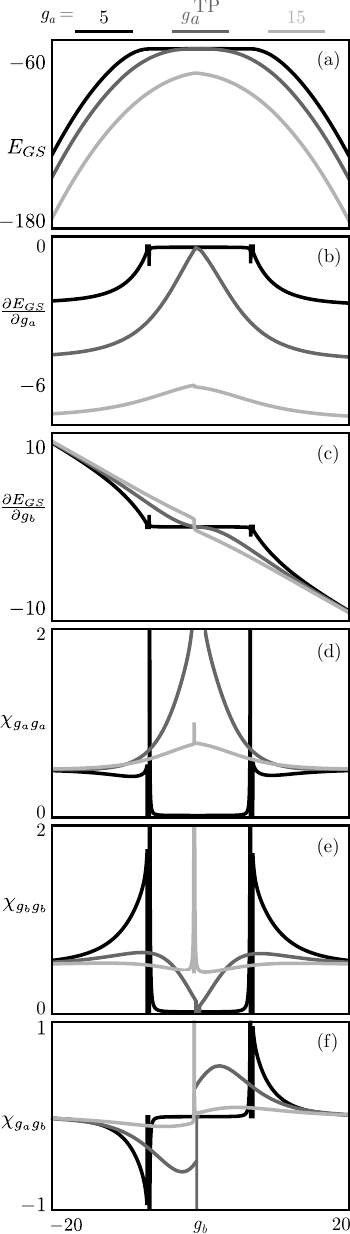}
	\caption{
		One-dimensional cuts of the ground-state energy and thermodynamic response functions.
		The three curves correspond to $g_a=5$,
		$g_a=g_a^{\rm TP}$, and $g_a=15$.
	}
	\label{fig:cut_response_ga_gb}
\end{figure}

\begin{figure}
	\centering
	\includegraphics{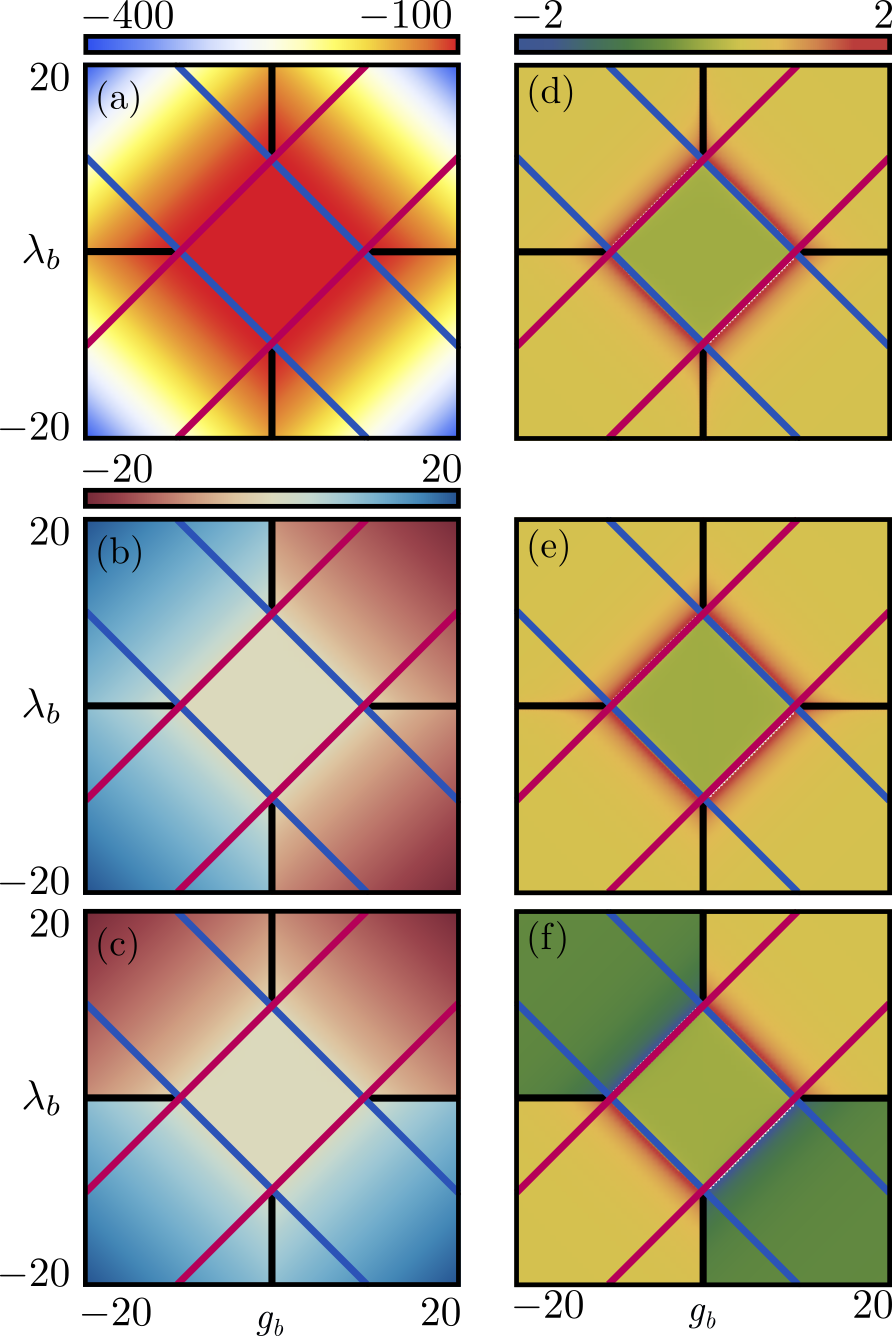}
	\caption{
		Ground-state energy and thermodynamic response functions in the $(g_b,\lambda_b)$ plane.
		(a) Ground-state energy $E_{\rm GS}$.
		(b) First derivative $\partial E_{\rm GS}/\partial g_b$.
		(c) First derivative $\partial E_{\rm GS}/\partial \lambda_b$.
		(d) Response function $\chi_{g_b g_b}$.
		(e) Response function $\chi_{\lambda_b \lambda_b}$.
		(f) Mixed response function $\chi_{g_b\lambda_b}$.
	}
	\label{fig:energy_response_gb_lambdab}
\end{figure}

\begin{figure}
	\centering
	\includegraphics{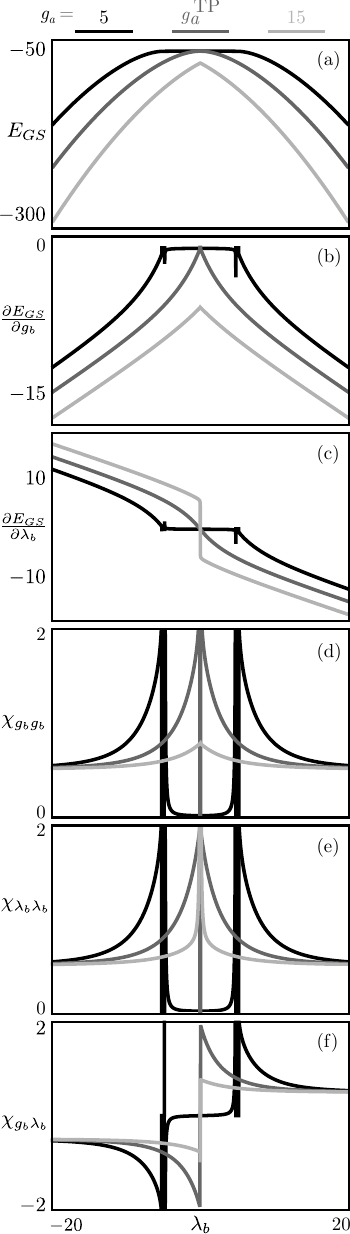}
	\caption{
		One-dimensional cuts of the ground-state energy and thermodynamic response functions.
		The three curves correspond to $g_b=5$,
		$g_b=g_b^{\rm TP}$, and $g_b=15$.
	}
	\label{fig:cut_response_gb_lambdab}
\end{figure}

Figure~\ref{fig:energy_response_ga_gb}(a) displays the quantum-corrected ground-state energy throughout the $(g_a,g_b)$ plane with
$\lambda_a=0.1$ and $\lambda_b=1.8$. The energy remains continuous across the entire phase diagram, including the triple points, indicating that the multicritical topology is not directly visible in the energy itself. Instead, its thermodynamic signatures emerge in the derivatives of the ground-state energy. Figs.~\ref{fig:energy_response_ga_gb}(b) and \ref{fig:energy_response_ga_gb}(c) present the first derivatives of the ground-state energy, $\partial E_{\rm GS}/\partial g_a$ and $\partial E_{\rm GS}/\partial g_b$. Across the SR$_r$--SR$_i$ first-order boundary, clear discontinuities are observed in Fig.~\ref{fig:energy_response_ga_gb}(c), whereas no discernible discontinuity is resolved in Fig.~\ref{fig:energy_response_ga_gb}(b).
This anisotropic response reflects the nearly horizontal orientation of the first-order boundary for
$\lambda_a=0.1$ and $\lambda_b=1.8$, such that the discontinuity of the ground-state-energy gradient is predominantly projected onto the $g_b$ direction.

The second derivatives of the ground-state energy,
$\chi_{AA}=-\frac{\partial^2E_{\rm GS}}{\partial g_a^2},
\chi_{BB}=-\frac{\partial^2E_{\rm GS}}{\partial g_b^2}, \chi_{AB}=-\frac{\partial^2E_{\rm GS}}{\partial g_a\partial g_b},$ are shown in Figs.~\ref{fig:energy_response_ga_gb}(d)--(f). These response functions exhibit pronounced enhancements along the continuous soft-mode boundaries, reflecting the rapid increase of quantum fluctuations as the lowest Bogoliubov excitation softens. Consequently, they provide a sensitive thermodynamic probe of the competing coordinate-like and momentum-like collective excitations. Although the detailed response patterns differ among the three susceptibility components because they probe the curvature of the ground-state energy along different directions in parameter space, they all faithfully reproduce the geometry of the continuous soft-mode boundaries identified in Fig.~\ref{fig:TwoTPsoftmode_maps}. Near the multicritical region, where the continuous phase boundaries intersect the first-order transition line, the response patterns undergo characteristic rearrangements associated with the competition between the coordinate-like and momentum-like collective excitations. This close correspondence establishes that the thermodynamic response is a direct manifestation of the underlying Bogoliubov instability structure.

Figure~\ref{fig:cut_response_ga_gb}(a) shows the ground-state energy along the three one-dimensional cuts taken at three representative values of $g_a$: $g_a=5$, $g_a=g_a^{\rm TP}$, and $g_a=15$. The ground-state energy remains continuous along all three cuts but exhibits different curvature near the phase boundaries.
The derivative with respect to $g_a$ in Figs.~\ref{fig:cut_response_ga_gb}(b)
 evolves from a smooth profile to a cusp-like structure at the triple-point cut, indicating the strongest sensitivity of the ground-state energy to parameter variations in the multicritical regime. By contrast, the derivative with respect to $g_b$ in Fig.~\ref{fig:cut_response_ga_gb}(c)
 develops kink-like discontinuities upon crossing the $\mathrm{SR}_{r}$--$\mathrm{SR}_{i}$
 first-order boundary, directly reflecting the branch-selection mechanism.
 The response functions in Figs.~\ref{fig:cut_response_ga_gb}(d)–(f) exhibit distinct evolutions for the three representative cuts. Away from the multicritical region $g_a=5$, the response enhancements associated with the two continuous phase boundaries remain well separated. At the triple-point cut, the two enhancements merge into a single multicritical structure, reflecting the simultaneous competition between the coordinate-like and momentum-like collective excitations. Deeper in the superradiant regime, at $g_a=15$, the response profiles become broader and smoother, while the mixed response changes sign across the first-order boundary, reflecting the switching between the competing SR-r and SR-i ground-state branches.

Together, Figs.~\ref{fig:TwoTPsoftmode_maps}--\ref{fig:cut_response_ga_gb} establish the correspondence between the Bogoliubov soft-mode structure and the thermodynamic signatures of the two-triple-point topology. Comparing Figs.~\ref{fig:TwoTPsoftmode_maps} and ~\ref{fig:energy_response_ga_gb}, the spatial organization of the thermodynamic responses in the ($g_a, g_b$) plane closely follows the distribution of the underlying soft modes. This correspondence is resolved more explicitly along the three representative cuts in Figs.~\ref{fig:TwoTPSoftModeCuts} and Figs.~\ref{fig:cut_response_ga_gb}. At  $g_a=5$, the two separated soft-mode instabilities produce distinct thermodynamic response features. At $g_a=g_a^{\rm TP}$, their simultaneous softening gives rise to a merged multicritical response structure. At $g_a=15$, the switching between the competing SR-r and SR-i branches across the first-order boundary is reflected in the discontinuities of the energy derivatives and the associated response profiles.

Figure~\ref{fig:energy_response_gb_lambdab}(a) shows the quantum-corrected ground-state energy in the $(g_b,\lambda_b)$ plane with $g_a=0.1$ and $\lambda_a=1.8$. As in the two-triple-point topology, the nergy remains continuous throughout the phase diagram and exhibits no singular feature at the four triple points. Instead, the multicritical structure emerges only in the derivatives of the ground-state energy.

The first derivatives shown in Figs.~\ref{fig:energy_response_gb_lambdab}(b) and
\ref{fig:energy_response_gb_lambdab}(c) provide a much clearer visualization of the multicritical topology. While both derivatives remain continuous across the continuous soft-mode boundaries, they exhibit distinct discontinuities across different segments of the first-order boundary. Specifically, $\partial E_{\rm GS}/\partial g_b$ develops a pronounced discontinuity along the vertical first-order boundary, whereas $\partial E_{\rm GS}/\partial \lambda_b$ exhibits the corresponding discontinuity along the horizontal first-order boundary. Together, these complementary response patterns partition the parameter plane into four symmetry-related sectors and directly reflect the redistribution of the competing coordinate-like and momentum-like collective excitations underlying the four-triple-point topology.
The second derivatives shown in
Figs.~\ref{fig:energy_response_gb_lambdab}(d)-(f)
reveal a response structure that differs qualitatively from
that of the two-triple-point topology. The two diagonal
components,
$\chi_{g_bg_b}=-\partial^2E_{\rm GS}/\partial g_b^2$
and
$\chi_{\lambda_b\lambda_b}
=-\partial^2E_{\rm GS}/\partial\lambda_b^2$,
are enhanced along the entire diamond-shaped continuous
boundary. In contrast to the two-triple-point case, the diagonal responses trace all
four symmetry-related boundary segments, although their intensities vary
along the diamond-shaped critical network. The mixed response,
$\chi_{g_b\lambda_b}
=-\partial^2 E_{\rm GS}/(\partial g_b\partial\lambda_b)$,
exhibits a qualitatively different, sign-alternating pattern across the
four parameter sectors, reflecting the directional correlation between
variations of $g_b$ and $\lambda_b$. Thus, while the diagonal responses
primarily resolve the closed critical network, the mixed response further
reveals the internal four-sector structure generated by the competition
between the coordinate and momentum collective excitations.

Figure.~\ref{fig:cut_response_gb_lambdab} presents representative one-dimensional cuts at $g_b=5$, $g_b=g_b^{\rm TP}$, and $g_b=15$. These cuts now probe the fourfold-symmetric diamond-shaped phase diagram, providing a one-dimensional view of how the thermodynamic response is reorganized by the higher multicritical symmetry.
Figure~\ref{fig:cut_response_gb_lambdab}(a) shows that the ground-state energy remains smooth for all three cuts, indicating that the multicritical topology is not directly encoded in the energy itself. Compared with Fig.~\ref{fig:cut_response_ga_gb}(a), the energy profiles exhibit a much higher degree of reflection symmetry about $\lambda_b=0$, inherited from the diamond-shaped organization of the phase diagram.
The first derivatives shown in Figs.~10(b) and~10(c) resolve the phase boundaries more clearly than the ground-state energy itself. Along the $g_b=5$ cut, the derivative $\partial E_{\rm GS}/\partial g_b$ develops a broad central plateau bounded by two symmetry-related continuous phase boundaries, reflecting the symmetric passage of the cut through the normal-phase region. At the triple-point cut, these two boundary features merge into a cusp-like structure. By contrast, $\partial E_{\rm GS}/\partial\lambda_b$ exhibits an approximately antisymmetric profile together with a pronounced discontinuity across the central first-order boundary, directly reflecting the switching between the competing SR-r and SR-i ground-state branches.
The second derivatives shown in Figs.~\ref{fig:cut_response_gb_lambdab}(d)-(f) reveal an even more pronounced distinction from the two-triple-point topology. The two diagonal susceptibilities,
$\chi_{g_b g_b}$ and $\chi_{\lambda_b\lambda_b}$,
differ from the anisotropic response profiles of
Fig.~\ref{fig:cut_response_ga_gb}
by displaying more reflection-symmetric critical structures that clearly resolve
all four symmetry-related continuous boundaries. In contrast, the mixed susceptibility $\chi_{g_b\lambda_b}$ retains a strongly sign-changing profile across the central first-order boundary, indicating that the directional competition between the coordinate-like and momentum-like collective excitations persists despite the enhanced spatial symmetry of the phase diagram.

Taken together, the one-dimensional cuts demonstrate that the transition from the two- to the four-triple-point topology is reflected not simply by the appearance of additional multicritical points, but by a qualitative reorganization of the thermodynamic response. While the two-triple-point topology concentrates the competition between the coordinate-like and momentum-like collective excitations around a pair of symmetry-related multicritical centers, the four-triple-point topology redistributes the same competition among four equivalent sectors. Consequently, the diagonal thermodynamic responses recover the enhanced symmetry of the diamond-shaped phase diagram, whereas the mixed response continues to encode the directional competition between the two collective excitation channels. 

\section{Conclusion}

We have shown that tunable multicritical topologies in a generalized two-mode quantum Rabi model originate from the competition between two Bogoliubov soft-mode channels associated with coordinate-like and momentum-like collective excitations. In the $U(1)$-symmetric limit, the system supports a continuously degenerate superradiant manifold accompanied by a gapless Goldstone-like collective mode. Finite anisotropy lifts this degeneracy and reorganizes the phase diagram into distinct multicritical topologies, including the two-triple-point and four-triple-point structures.
By developing a unified Bogoliubov theory, we demonstrated that the redistribution of the competing soft modes determines both the geometry of the phase boundaries and the arrangement of the triple points. The collective soft-mode softening exactly reproduces the mean-field phase boundaries, thereby establishing a unified description of symmetry breaking, collective excitations, and multicritical phase topology. Furthermore, we showed that the same soft-mode competition is directly encoded in the quantum ground-state energy and its thermodynamic response functions, providing observable signatures of the underlying instability mechanism.
Our results establish a direct correspondence between Bogoliubov soft modes, multicritical topology, and thermodynamic observables, identifying competing collective excitations as the microscopic origin of tunable multicriticality in generalized quantum Rabi systems. More broadly, the framework developed here provides a unified perspective connecting symmetry, collective fluctuations, phase topology, and thermodynamic responses in strongly coupled light--matter systems. We anticipate that these ideas can be extended to a broad range of multimode cavity-QED and quantum-optical platforms, opening new opportunities for engineering multicritical quantum phases through the controlled competition of collective excitation channels.

\begin{acknowledgments}
This work was supported by the China Scholarship Council (CSC) 
under Scholarship No.~202406310190.
\end{acknowledgments}


\bibliography{references}

@article{Dicke1954,
	author = {R. H. Dicke},
	title = {Coherence in Spontaneous Radiation Processes},
	journal = {Physical Review},
	volume = {93},
	number = {1},
	pages = {99--110},
	year = {1954},
	doi = {10.1103/PhysRev.93.99}
}

@article{Hepp1973,
	author = {Klaus Hepp and Elliott H. Lieb},
	title = {On the Superradiant Phase Transition for Molecules in a Quantized Radiation Field: the Dicke Maser Model},
	journal = {Annals of Physics},
	volume = {76},
	number = {2},
	pages = {360--404},
	year = {1973},
	doi = {10.1016/0003-4916(73)90039-0}
}

@article{Wang1973,
	author = {Y. K. Wang and F. T. Hioe},
	title = {Phase Transition in the Dicke Model of Superradiance},
	journal = {Physical Review A},
	volume = {7},
	number = {3},
	pages = {831--836},
	year = {1973},
	doi = {10.1103/PhysRevA.7.831}
}

@article{Emary2003a,
	author = {Clive Emary and Tobias Brandes},
	title = {Chaos and the Quantum Phase Transition in the Dicke Model},
	journal = {Physical Review E},
	volume = {67},
	pages = {066203},
	year = {2003},
	doi = {10.1103/PhysRevE.67.066203}
}

@article{Emary2003b,
	author = {Clive Emary and Tobias Brandes},
	title = {Quantum Chaos Triggered by Precursors of a Quantum Phase Transition},
	journal = {Physical Review Letters},
	volume = {90},
	pages = {044101},
	year = {2003},
	doi = {10.1103/PhysRevLett.90.044101}
}

@article{Braak2011,
	author = {Daniel Braak},
	title = {Integrability of the Rabi Model},
	journal = {Physical Review Letters},
	volume = {107},
	pages = {100401},
	year = {2011},
	doi = {10.1103/PhysRevLett.107.100401}
}

@article{Hwang2015,
	author = {Myung-Joong Hwang and Ricardo Puebla and Martin B. Plenio},
	title = {Quantum Phase Transition and Universal Dynamics in the Rabi Model},
	journal = {Phys. Rev. Lett.},
	volume = {115},
	pages = {180404},
	year = {2015},
	doi = {10.1103/PhysRevLett.115.180404}
}

@article{Liu2017,
	author = {Maoxin Liu and Stefano Chesi and Zu-Jian Ying and Xiaosong Chen and Hong-Gang Luo and Hai-Qing Lin},
	title = {Universal Scaling and Critical Exponents of the Anisotropic Quantum Rabi Model},
	journal = {Phys. Rev. Lett.},
	volume = {119},
	pages = {220601},
	year = {2017},
	doi = {10.1103/PhysRevLett.119.220601}
}

@article{Hwang2018,
	author = {Myung-Joong Hwang and Peter Rabl and Martin B. Plenio},
	title = {Dissipative Phase Transition in the Open Quantum Rabi Model},
	journal = {Phys. Rev. A},
	volume = {97},
	pages = {013825},
	year = {2018},
	doi = {10.1103/PhysRevA.97.013825}
}

@article{Chen2021,
	author = {Xiang-You Chen and Liwei Duan and Daniel Braak and Qing-Hu Chen},
	title = {Multiple Ground-State Instabilities in the Anisotropic Quantum Rabi Model},
	journal = {Phys. Rev. A},
	volume = {103},
	pages = {043708},
	year = {2021},
	doi = {10.1103/PhysRevA.103.043708}
}

@article{Ying2021,
	author = {Zu-Jian Ying},
	title = {Symmetry-Breaking Patterns, Tricriticalities, and Quadruple Points in the Quantum Rabi Model with Bias and Nonlinear Interaction},
	journal = {Phys. Rev. A},
	volume = {103},
	pages = {063701},
	year = {2021},
	doi = {10.1103/PhysRevA.103.063701}
}

@article{RabiTriangle2021,
	author = {Yu-Yu Zhang and Zi-Xiang Hu and Libin Fu and Hong-Gang Luo and Han Pu and Xue-Feng Zhang},
	title = {Quantum Phases in a Quantum Rabi Triangle},
	journal = {Phys. Rev. Lett.},
	volume = {127},
	pages = {063602},
	year = {2021},
	doi = {10.1103/PhysRevLett.127.063602}
}

@article{Lyu2024,
	author = {Guitao Lyu and Korbinian Kottmann and Martin B. Plenio and Myung-Joong Hwang},
	title = {Multicritical Dissipative Phase Transitions in the Anisotropic Open Quantum Rabi Model},
	journal = {Phys. Rev. Research},
	volume = {6},
	pages = {033075},
	year = {2024},
	doi = {10.1103/PhysRevResearch.6.033075}
}

@article{Wu2024,
	author = {Ze Wu and Changsheng Hu and Tianyun Wang and Yuquan Chen and Yuchen Li and Liqiang Zhao and Xin-You Lu and Xinhua Peng},
	title = {Experimental Quantum Simulation of Multicriticality in Closed and Open Rabi Model},
	journal = {Phys. Rev. Lett.},
	volume = {133},
	pages = {173602},
	year = {2024},
	doi = {10.1103/PhysRevLett.133.173602}
}

@article{Xu2024,
	author = {Yilun Xu and Feng-Xiao Sun and Wei Zhang and Qiongyi He and Han Pu},
	title = {Phase Transition and Multistability in Dicke Dimer},
	journal = {Phys. Rev. Lett.},
	volume = {133},
	pages = {233604},
	year = {2024},
	doi = {10.1103/PhysRevLett.133.233604}
}

@article{Zhu2024,
	author = {Gui-Lei Zhu and Chang-Sheng Hu and Hui Wang and Wei Qin and Xin-You Lu and Franco Nori},
	title = {Nonreciprocal Superradiant Phase Transitions and Multicriticality in a Cavity QED System},
	journal = {Phys. Rev. Lett.},
	volume = {132},
	pages = {193602},
	year = {2024},
	doi = {10.1103/PhysRevLett.132.193602}
}

@article{Adiv2024,
	author = {Ofri Adiv and Scott Parkins},
	title = {Dynamics of a Generalized Dicke Model for Spin-1 Atoms},
	journal = {Phys. Rev. A},
	volume = {110},
	pages = {023705},
	year = {2024},
	doi = {10.1103/PhysRevA.110.023705}
}

@article{Xie2025,
	author = {Xuan Xie and Cheng Liu and Lin-Lin Jiang and Jin-Feng Huang},
	title = {Controlling the Superradiant Phase Transition in the Quantum Rabi Model},
	journal = {Phys. Rev. A},
	volume = {111},
	pages = {023708},
	year = {2025},
	doi = {10.1103/PhysRevA.111.023708}
}

@article{Friedemann2018,
	author = {Sven Friedemann and Will J. Duncan and Max Hirschberger and Thomas W. Bauer and Robert Kuechler and Andreas Neubauer and Manuel Brando and Christian Pfleiderer and F. Malte Grosche},
	title = {Quantum Tricritical Points in NbFe2},
	journal = {Nature Physics},
	volume = {14},
	pages = {62--68},
	year = {2018},
	doi = {10.1038/nphys4242}
}


\clearpage

\clearpage
\onecolumngrid

\begin{center}
	{\large\bfseries
		Supplementary Material for\\[4pt]
		Competing Soft Modes and Tunable Multicriticality in a\\
		Generalized Two-Mode Quantum Rabi Model
	}
	
	\vspace{0.8em}
	
	Xiufeng Cao,\quad Ofri Adiv,\quad and Neil G. R. Broderick
\end{center}

\vspace{1.5em}

\appendix

\makeatletter
\@addtoreset{equation}{section}
\@addtoreset{figure}{section}
\@addtoreset{table}{section}
\makeatother

\renewcommand{\theequation}{\thesection\arabic{equation}}
\renewcommand{\thefigure}{\thesection\arabic{figure}}
\renewcommand{\thetable}{\thesection\arabic{table}}

\section{Mean-Field Steady-State Solutions}
	\label{app:mean-field}
To derive the semiclassical steady-state solutions used in the main text, we start from the Heisenberg equations associated with Eq.~\eqref{eq:Hamiltonian},
\begin{align}
	\dot a &= -i\omega_a a-\frac{i}{2}(g_a+\lambda_a)\sigma_x
	+\frac{1}{2}(g_a-\lambda_a)\sigma_y,\\
	\dot b &= -i\omega_b b-\frac{i}{2}(g_b+\lambda_b)\sigma_x
	+\frac{1}{2}(g_b-\lambda_b)\sigma_y,\\
	\dot\sigma_x &= -\Delta_z\sigma_y
	+i(g_a-\lambda_a)(a-a^\dagger)\sigma_z \\
	&\quad
	+i(g_b-\lambda_b)(b-b^\dagger)\sigma_z, \nonumber \\
	\dot\sigma_y &= \Delta_z\sigma_x
	-(g_a+\lambda_a)(a+a^\dagger)\sigma_z \\
	&\quad
	-(g_b+\lambda_b)(b+b^\dagger)\sigma_z, \nonumber \\
	\dot{\sigma}_z &=
	(g_a+\lambda_a)(a+a^\dagger)\sigma_y
	-i(g_a-\lambda_a)(a-a^\dagger)\sigma_x\\
	&\quad
	+(g_b+\lambda_b)(b+b^\dagger)\sigma_y
	-i(g_b-\lambda_b)(b-b^\dagger)\sigma_x \nonumber .
\end{align}
Within the mean-field framework, we introduce the cavity-field order parameters
$\alpha=\langle a\rangle,\qquad \beta=\langle b\rangle $
and the collective spin expectation values
$X=\langle\sigma_x\rangle,\qquad
Y=\langle\sigma_y\rangle,\qquad
Z=\langle\sigma_z\rangle$.
Applying the standard mean-field decoupling approximation,
$
\langle a\sigma_z\rangle \approx \langle a\rangle\langle\sigma_z\rangle=\alpha Z,
$
and similarly for all higher-order operator products, the equations of motion reduce to the closed set of semiclassical equations:
\begin{align}
	\dot{\alpha} &= -i\omega_a \alpha - \frac{i}{2}A_+ X + \frac{1}{2}A_- Y, \\
	\dot{\beta} &= -i\omega_b \beta - \frac{i}{2}B_+ X + \frac{1}{2}B_- Y, \\
	\dot{X} &= -\Delta_z Y + iA_- (\alpha - \alpha^*) Z + iB_- (\beta - \beta^*) Z, \\
	\dot{Y} &= \Delta_z X - A_+ (\alpha + \alpha^*) Z - B_+ (\beta + \beta^*) Z,
\end{align}
with the definition $	A_\pm=g_a\pm\lambda_a$,
$B_\pm=g_b\pm\lambda_b$.The bosonic displacements are written as
\(
\alpha=\alpha_r+i\alpha_i
\)
and
\(
\beta=\beta_r+i\beta_i
\),
the steady-state conditions
\(
\dot{\alpha}=\dot{\beta}=0
\),
together with the spin conservation constraint
\(
X^2+Y^2+Z^2=1
\),
yield
\begin{align}
	\alpha_r&=-\frac{A_+}{2\omega_a}X,
	&
	\alpha_i&=-\frac{A_-}{2\omega_a}Y,
	\nonumber\\
	\beta_r&=-\frac{B_+}{2\omega_b}X,
	&
	\beta_i&=-\frac{B_-}{2\omega_b}Y.
	\label{eq:alpha_beta_sol}
\end{align}
Substituting these expressions into the spin equations gives
\begin{equation}
	Y
	\left[
	-\Delta_z
	+\mathcal{C}_{r}
	\right]=0,
\end{equation}
and
\begin{equation}
	X
	\left[
	\Delta_z
	-
\mathcal{C}_{i}
	\right]=0,
\end{equation}
with the definition $\mathcal{C}_{r}
=
\frac{A_{+}^{2}}{\omega_a}
+
\frac{B_{+}^{2}}{\omega_b}$,
$\mathcal{C}_{i}
=
\frac{A_{-}^{2}}{\omega_a}
+
\frac{B_{-}^{2}}{\omega_b}$. These equations separate the superradiant steady-state solutions into two distinct branches, characterized by real and imaginary bosonic displacements, respectively.

For the real-displacement branch $(X\neq0,\ Y=0)$, one obtains
\begin{equation}
	Z
	=
	-\frac{\Delta_z}{
		\mathcal{C}_{r}},
\end{equation}
and
\begin{equation}
	X
	=
	\pm
	\sqrt{
		1-
		\frac{
			\Delta_z^2
		}{\mathcal{C}_{r}^2	}
	}.
\end{equation}
This branch exists for $\mathcal{C}_{r}
\ge |\Delta_z|$ ,
with purely real bosonic displacements
\begin{equation}
	\alpha
	=
	\pm
	\frac{A_+}{2\omega_a}
	\sqrt{
		1-
		\frac{
			\Delta_z^2
		}{\mathcal{C}_{r}^2
	}},
\end{equation}

\begin{equation}
	\beta
	=
	\pm
	\frac{B_+}{2\omega_b}
	\sqrt{
		1-
		\frac{
			\Delta_z^2
		}{\mathcal{C}_{r}^2
	}}.
\end{equation}
Similarly, for the imaginary-displacement branch $(X=0,\ Y\neq0)$,
\begin{equation}
	Z
	=
	\frac{\Delta_z}{
		\mathcal{C}_{i}},
\end{equation}
and
\begin{equation}
	Y
	=
	\pm
	\sqrt{
		1-
		\frac{
			\Delta_z^2
		}{
			\mathcal{C}_{i}^2
		}
	}.
\end{equation}
This branch exists for $\mathcal{C}_{i}
\ge |\Delta_z|$,
with purely imaginary order parameters
\begin{equation}
	\alpha
	=
	\pm i
	\frac{A_-}{2\omega_a}
	\sqrt{
		1-
		\frac{
			\Delta_z^2
		}{\mathcal{C}_{i}^2
	}},
\end{equation}

\begin{equation}
	\beta
	=
	\pm i
	\frac{B_-}{2\omega_b}
	\sqrt{
		1-
		\frac{
			\Delta_z^2
		}{\mathcal{C}_{i}^2
	}}.
\end{equation}
Since $\Delta_z>0$ throughout this work, the absolute-value notation can be omitted. The normal-to-superradiant critical conditions therefore reduce to
$\mathcal{C}_{r}=\Delta_z$ and $\mathcal{C}_{i}=\Delta_z$
for the real- and imaginary-displacement branches, respectively. Physically, $\mathcal{C}_r$ and $\mathcal{C}_i$ define the
coordinate-like and momentum-like instability channels, respectively,
each becoming critical when the corresponding effective coupling reaches
$\Delta_z$. As shown below, these conditions coincide exactly with the
Bogoliubov soft-mode instabilities, directly linking the mean-field
bifurcation to collective-mode softening.

\section{Schrieffer-Wolff Transformation and Effective low-energy Hamiltonian in the normal phase}
	\label{app:sw-normalphase}
In this Appendix, we derive the low-energy effective bosonic Hamiltonian
underlying the Bogoliubov analysis in the main text. In the
large-frequency-ratio regime, the atomic excited state is energetically
separated from the low-energy bosonic sector and contributes primarily
through virtual transitions. We therefore employ a SW transformation to perturbatively eliminate the high-energy atomic
sector while retaining its virtual effects as effective interactions
among the bosonic modes. Starting from the Hamiltonian in
Eq.~\eqref{eq:Hamiltonian}, we carry out the SW transformation to second
order in the light--matter couplings and project the resulting Hamiltonian
onto the atomic ground-state manifold.
Expressing the bosonic operators in terms of the canonical quadratures
\(x_\mu=(\mu+\mu^\dagger)/\sqrt{2}\)
and
\(p_\mu=i(\mu^\dagger-\mu)/\sqrt{2}\)
with \(\mu=a,b\),
the Hamiltonian in Eq.~(\ref{eq:Hamiltonian}) becomes
\begin{equation}
	\begin{aligned}
		H
		&=
		\frac{\omega_a}{2}\left(x_a^2+p_a^2-1\right)
		+
		\frac{\omega_b}{2}\left(x_b^2+p_b^2-1\right)
		+
		\frac{\Delta_z}{2}\sigma_z
		\\
		&\quad
		+
		\frac{A_+}{\sqrt{2}}\sigma_x x_a
		-
		\frac{A_-}{\sqrt{2}}\sigma_y p_a
		\\
		&\quad
		+
		\frac{B_+}{\sqrt{2}}\sigma_x x_b
		-
		\frac{B_-}{\sqrt{2}}\sigma_y p_b.
	\end{aligned}
	\label{eq:Hxp}
\end{equation}
For the normal-phase branch (denoted by the subscript \(n\)),
we introduce the collective quadrature operators
\(X_n=(A_+ x_a+B_+ x_b)/\sqrt{2}\)
and
\(P_n=-(A_- p_a+B_- p_b)/\sqrt{2}\),
such that the interaction term takes the compact form
\(V_n=\sigma_x X_n+\sigma_y P_n\).
Following a SW transformation, we introduce the generator
\begin{equation}
	S_n
	=
	\frac{1}{\Delta_z}\sigma_y X_n
	-
	\frac{1}{\Delta_z}\sigma_x P_n ,
\end{equation}
and the Hamiltonian becomes
\(H_{\mathrm{n}}'
=
e^{iS_n} H e^{-iS_n}\).

Keeping terms up to second order, we obtain
\begin{equation}
	\begin{aligned}
		H_{\mathrm{n}}'
		\simeq&
		\frac{\omega_a}{2}(x_a^2+p_a^2-1)
		+
		\frac{\omega_b}{2}(x_b^2+p_b^2-1)
		+
		\frac{\Delta_z}{2}\sigma_z
		\\
		&\quad
		+
		\frac{\sigma_z}{2\Delta_z}
		\left(
		X_n^2
		+
		P_n^2
		\right)	.
	\end{aligned}
	\label{eq:HSRr_second}
\end{equation}
Projecting onto the low-energy spin sector \(\sigma_z=-1\), we finally obtain the effective bosonic Hamiltonian
\begin{equation}
	\begin{aligned}
		H_{\mathrm{n}}'
		=&-
		\frac{\Delta_z}{2}
		+
		\frac{\omega_a}{2}(x_a^2+p_a^2-1)
		+
		\frac{\omega_b}{2}(x_b^2+p_b^2-1)
		\\
		&\quad
		-
		\frac{1}{2\Delta_z}
		\left(
		A_+ x_a
		+
		B_+ x_b
		\right)^2
		\\
		&\quad
		-
		\frac{1}{2\Delta_z}
		\left(
		A_- p_a
		+
		B_- p_b
		\right)^2 .
	\end{aligned}
	\label{eq:HeffNP}
\end{equation}

To explicitly resolve the contributions of the rotating- and
counter-rotating-wave couplings to the low-energy bosonic interactions,
we rewrite the normal-phase effective Hamiltonian in
Eq.~\eqref{eq:HeffNP} in terms of bosonic creation and annihilation
operators as
\begin{align}
\hat H_{\rm eff}^{N}
&=
\tilde E_N
+
\omega_a^{\rm eff} a^\dagger a
+
\omega_b^{\rm eff} b^\dagger b
\nonumber\\
&\quad
+
\eta_a\left(a^2+a^{\dagger 2}\right)
+
\eta_b\left(b^2+b^{\dagger 2}\right)
\nonumber\\
&\quad
+
J_{\rm eff}\left(a^\dagger b+a b^\dagger\right)
+
\xi_{\rm eff}\left(a b+a^\dagger b^\dagger\right).
\label{eq:Heff_ab_simple}
\end{align}
Here, the renormalized mode frequencies are
\begin{equation}
\omega_a^{\rm eff}
=
\omega_a
-
\frac{g_a^2+\lambda_a^2}{\Delta_z},
\qquad
\omega_b^{\rm eff}
=
\omega_b
-
\frac{g_b^2+\lambda_b^2}{\Delta_z},
\label{eq:omegaeff_simple}
\end{equation}
the single-mode anomalous couplings are
\begin{equation}
\eta_a
=
-\frac{g_a\lambda_a}{\Delta_z},
\qquad
\eta_b
=
-\frac{g_b\lambda_b}{\Delta_z},
\label{eq:etaeff_simple}
\end{equation}
and the intermode couplings are
\begin{equation}
J_{\rm eff}
=
-\frac{g_a g_b+\lambda_a\lambda_b}{\Delta_z},
\qquad
\xi_{\rm eff}
=
-\frac{g_a\lambda_b+\lambda_a g_b}{\Delta_z}.
\label{eq:Jeff_xieff_simple}
\end{equation}
The constant contribution is
\begin{equation}
\tilde E_N
=
-\frac{\Delta_z}{2}
-
\frac{
g_a^2+\lambda_a^2+g_b^2+\lambda_b^2
}{
2\Delta_z
}.
\label{eq:EtildeN}
\end{equation}
This representation explicitly separates the number-conserving and
anomalous contributions generated by the light--matter couplings.
In particular, the anomalous terms arise from second-order processes
involving both rotating- and counter-rotating-wave couplings:
$\eta_a$ and $\eta_b$ describe single-mode squeezing, whereas
$\xi_{\rm eff}$ describes two-mode pairing. These terms vanish in the
rotating-wave limit and induce the particle--hole mixing underlying
the BdG excitation spectrum and the associated soft-mode instabilities.

\section{Soft-Mode Condition and Ground-State Energy in the Normal Phase}
	\label{app:sm-normalphase}
	
To characterize the collective excitations of the normal phase, we diagonalize
the quadratic effective Hamiltonian derived above. Introducing the Nambu spinor $\Psi=
\begin{pmatrix}
a & b & a^\dagger & b^\dagger
\end{pmatrix}^{T}$,
Eq.~\eqref{eq:Heff_ab_simple} can be written in the standard bosonic
BdG form
\begin{equation}
\hat{H}_{\rm eff}^{N}
=
\tilde{E}_{N}
+
\frac{1}{2}
\Psi^\dagger
\mathcal{H}_{\rm BdG}^{N}
\Psi
-
\frac{1}{2}{\rm Tr}(h_N),
\label{eq:HBdG}
\end{equation}
where
\begin{equation}
\mathcal{H}_{\rm BdG}^{N}
=
\begin{pmatrix}
h_N & \Delta_N\\
\Delta_N & h_N
\end{pmatrix},
\qquad
h_N=
\begin{pmatrix}
\omega_a^{\rm eff} & J_{\rm eff}\\
J_{\rm eff} & \omega_b^{\rm eff}
\end{pmatrix},
\qquad
\Delta_N=
\begin{pmatrix}
2\eta_a & \xi_{\rm eff}\\
\xi_{\rm eff} & 2\eta_b
\end{pmatrix}.
\label{eq:BdGblocks}
\end{equation}
Here, $h_N$ contains the number-conserving terms, including the renormalized
mode frequencies and intermode hopping, whereas $\Delta_N$ collects the
anomalous terms associated with single-mode squeezing and two-mode pairing.
The latter generate the particle--hole mixing characteristic of the bosonic
BdG problem.

Because the Bogoliubov transformation must preserve the bosonic
commutation relations, the collective excitation frequencies are
determined by the generalized BdG eigenvalue problem
\begin{equation}
\Sigma_z\mathcal{H}_{\rm BdG}^{N}\Phi_n
=
\Omega_n\Phi_n,
\qquad
\Sigma_z=
\begin{pmatrix}
I_2 & 0\\
0 & -I_2
\end{pmatrix},
\label{eq:BdGEigenN}
\end{equation}
where $\Sigma_z$ is the metric in Nambu (particle--hole) space.
The two positive eigenvalues define the physical excitation branches
$\Omega_{N,\pm}$. For the present two-mode system, their squared
frequencies are
\begin{equation}
\Omega_{N,\pm}^{2}
=
\frac{1}{2}
\left[
T_N
\pm
\sqrt{
T_N^2-4D_N
}
\right],
\label{eq:Omega2}
\end{equation}
with
\begin{equation}
T_N
=
{\rm Tr}\!\left(K_p^N K_x^N\right),
\qquad
D_N
=
\det\!\left(K_p^N K_x^N\right).
\label{eq:TDN}
\end{equation}
The coordinate-sector stiffness matrix is
\begin{equation}
K_x^N
=
h_N+\Delta_N
=
\begin{pmatrix}
\displaystyle
\omega_a-\frac{A_+^2}{\Delta_z}
&
\displaystyle
-\frac{A_+B_+}{\Delta_z}
\\[2mm]
\displaystyle
-\frac{A_+B_+}{\Delta_z}
&
\displaystyle
\omega_b-\frac{B_+^2}{\Delta_z}
\end{pmatrix},
\label{eq:KxN}
\end{equation}
while the momentum-sector stiffness matrix is
\begin{equation}
K_p^N
=
h_N-\Delta_N
=
\begin{pmatrix}
\displaystyle
\omega_a-\frac{A_-^2}{\Delta_z}
&
\displaystyle
-\frac{A_-B_-}{\Delta_z}
\\[2mm]
\displaystyle
-\frac{A_-B_-}{\Delta_z}
&
\displaystyle
\omega_b-\frac{B_-^2}{\Delta_z}
\end{pmatrix}.
\label{eq:KpN}
\end{equation}
Thus, $K_x^N$ and $K_p^N$ describe the coordinate-like and
momentum-like collective fluctuation sectors, respectively.

The stability of the normal phase is governed by the lower branch
$\Omega_{N,-}$. From Eq.~\eqref{eq:Omega2}, the soft-mode condition
$\Omega_{N,-}=0$ is equivalent to
\begin{equation}
D_N
=
\det\!\left(K_p^N K_x^N\right)
=
\det K_p^N\,\det K_x^N
=
0.
\label{eq:softconditionN}
\end{equation}
Hence, the normal phase can lose stability through either $\det K_x^N=0$
or $det K_p^N=0$.
Evaluating the two determinant conditions yields $\mathcal C_r=\Delta_z$, $\mathcal C_i=\Delta_z$,
corresponding to the coordinate-like and momentum-like softening,
respectively. These conditions coincide exactly with the mean-field
critical boundaries of the $N$--$\mathrm{SR}_r$ and
$N$--$\mathrm{SR}_i$ transitions, establishing the equivalence between
the mean-field and BdG descriptions of the normal-phase instabilities.

Having obtained the Bogoliubov excitation spectrum, we now evaluate
the corresponding quantum correction to the normal-phase ground-state
energy. After the Bogoliubov transformation, the effective Hamiltonian
takes the diagonal form,
\begin{equation}
E_{\rm GS}^{N}
=
\tilde{E}_{N}
-
\frac{1}{2}{\rm Tr}(h_N)
+
\frac{1}{2}
\left(
\Omega_+
+
\Omega_-
\right).
\label{eq:EGS_BdG}
\end{equation}
Here $-\frac{1}{2}{\rm Tr}(h_N)$ is the constant contribution generated
when the quadratic Hamiltonian is written in Nambu form, while
$\frac{1}{2}(\Omega_++\Omega_-)$ is the zero-point energy of the two
Bogoliubov quasiparticle modes. Using
\begin{equation}
\tilde{E}_{N}
=
-\frac{\Delta_z}{2}
-
\frac{
g_a^2+\lambda_a^2+g_b^2+\lambda_b^2
}{
2\Delta_z
}
\end{equation}
together with
\begin{equation}
{\rm Tr}(h_N)
=
\omega_a+\omega_b
-
\frac{
g_a^2+\lambda_a^2+g_b^2+\lambda_b^2
}{
\Delta_z
},
\end{equation}
the coupling-dependent constant contributions cancel exactly. The
ground-state energy therefore becomes
\begin{equation}
E_{\rm GS}^{N}
=
\bar{E}_{N}
+
\frac{1}{2}
\left(
\Omega_+
+
\Omega_-
-
\omega_a
-
\omega_b
\right),
\label{eq:EGS_final}
\end{equation}
where $\bar{E}_{N}
=
-\frac{\Delta_z}{2}$
is the normal-phase mean-field energy. The second term in
Eq.~\eqref{eq:EGS_final} is the zero-point-energy correction generated by
the collective Bogoliubov fluctuations, including mode hybridization and
squeezing. Its nonanalytic behavior as
$\Omega_{N,-}\to0$ reflects the closing of the
collective excitation gap at the normal-to-superradiant critical boundary.

\section{Unified Phase-Locking Description of Superradiant Ordering}
	\label{app:unified-phaselocking}

To describe the two superradiant branches within a common framework,
we introduce the coherent displacements  $a=\alpha+\hat c$, $b=\beta+\hat d$, where $\alpha$ and $\beta$ denote the condensate amplitudes, while
$\hat c$ and $\hat d$ describe quantum fluctuations around the
displaced state. 
Substituting displacements into the original Hamiltonian
gives
\begin{align}
\hat H_{\rm disp}
=&\,
\omega_a|\alpha|^2
+\omega_b|\beta|^2
+\omega_a\hat c^\dagger\hat c
+\omega_b\hat d^\dagger\hat d
+\frac{\Delta_z}{2}\hat\sigma_z
\nonumber\\
&+
h_+\hat\sigma_+
+h_+^*\hat\sigma_-
\nonumber\\
&+
g_a
\left(
\hat c\hat\sigma_+
+\hat c^\dagger\hat\sigma_-
\right)
+
\lambda_a
\left(
\hat c\hat\sigma_-
+\hat c^\dagger\hat\sigma_+
\right)
\nonumber\\
&+
g_b
\left(
\hat d\hat\sigma_+
+\hat d^\dagger\hat\sigma_-
\right)
+
\lambda_b
\left(
\hat d\hat\sigma_-
+\hat d^\dagger\hat\sigma_+
\right),
\label{eq:Hdisp}
\end{align}
where $h_+
=
g_a\alpha+\lambda_a\alpha^*
+
g_b\beta+\lambda_b\beta^*$
is the coherent transverse field generated by the condensates.
The displacement-dependent spin sector becomes
\begin{equation}
\hat H_{\rm spin}^{\rm disp}
=
\frac{\Delta_z}{2}\hat\sigma_z
+
h_+\hat\sigma_+
+
h_+^*\hat\sigma_-.
\label{eq:Hspin_disp}
\end{equation}
Writing $h_+=|h_+|e^{-i\phi}$,
the spin quantization axis can be rotated along the corresponding
effective field, giving
\begin{equation}
\hat H_{\rm spin}^{\rm rot}
=
\frac{\Delta_z'}{2}\hat\tau_z,
\qquad
\Delta_z'
=
\sqrt{\Delta_z^2+4|h_+|^2},
\label{eq:Hspin_rot}
\end{equation}
with
\begin{equation}
\cos\theta
=
\frac{\Delta_z}{\Delta_z'},
\qquad
\sin\theta
=
\frac{2|h_+|}{\Delta_z'}.
\label{eq:rotation_angle}
\end{equation}
In the dressed-spin ground state, the transverse spin expectation
values are
\begin{equation}
\langle\hat\sigma_+\rangle
=
-\frac{\sin\theta}{2}e^{i\phi},
\qquad
\langle\hat\sigma_-\rangle
=
-\frac{\sin\theta}{2}e^{-i\phi}.
\label{eq:spin_expectation}
\end{equation}

Requiring the linear fluctuation terms in $\hat c$ and $\hat d$ to vanish
gives the self-consistency conditions for the condensate amplitudes,
\begin{equation}
\alpha
=
-\frac{1}{\omega_a}
\left(
g_a\langle\sigma_-\rangle
+
\lambda_a\langle\sigma_+\rangle
\right),
\qquad
\beta
=
-\frac{1}{\omega_b}
\left(
g_b\langle\sigma_-\rangle
+
\lambda_b\langle\sigma_+\rangle
\right).
\label{eq:self_consistency_alpha_beta}
\end{equation}
Using the dressed-spin expectation values in Eq.~\eqref{eq:spin_expectation},
these relations become
\begin{equation}
\alpha
=
\frac{\sin\theta}{2\omega_a}
\left(
g_a e^{-i\phi}
+
\lambda_a e^{i\phi}
\right),
\qquad
\beta
=
\frac{\sin\theta}{2\omega_b}
\left(
g_b e^{-i\phi}
+
\lambda_b e^{i\phi}
\right).
\label{eq:alpha_beta_phase}
\end{equation}
Substituting Eq.~\eqref{eq:alpha_beta_phase} into Eq.~\eqref{eq:hplus}
gives
\begin{equation}
\begin{aligned}
h_+
=
\frac{\sin\theta}{2}
\Bigg[
&
\left(
\frac{g_a^2+\lambda_a^2}{\omega_a}
+
\frac{g_b^2+\lambda_b^2}{\omega_b}
\right)e^{-i\phi}
\\
&+
\left(
\frac{2g_a\lambda_a}{\omega_a}
+
\frac{2g_b\lambda_b}{\omega_b}
\right)e^{i\phi}
\Bigg].
\end{aligned}
\label{eq:hplus_phase}
\end{equation}
Using
\begin{equation}
\frac{\mathcal C_r+\mathcal C_i}{2}
=
\frac{g_a^2+\lambda_a^2}{\omega_a}
+
\frac{g_b^2+\lambda_b^2}{\omega_b},
\qquad
\frac{\mathcal C_r-\mathcal C_i}{2}
=
\frac{2g_a\lambda_a}{\omega_a}
+
\frac{2g_b\lambda_b}{\omega_b},
\end{equation}
Eq.~\eqref{eq:hplus_phase} can be rewritten as
\begin{equation}
h_+
=
\frac{\sin\theta}{2}
\left[
\frac{\mathcal C_r+\mathcal C_i}{2}e^{-i\phi}
+
\frac{\mathcal C_r-\mathcal C_i}{2}e^{i\phi}
\right].
\label{eq:hplus_collective}
\end{equation}

Finally, using $h_+=|h_+|e^{-i\phi}$ together with
$\sin\theta=2|h_+|/\Delta_z'$, Eq.~\eqref{eq:hplus_collective}
reduces to
\begin{equation}
\Delta_z'
=
\frac{\mathcal C_r+\mathcal C_i}{2}
+
\frac{\mathcal C_r-\mathcal C_i}{2}e^{2i\phi},
\label{eq:phase_locking}
\end{equation}
which is the phase-locking equation used in the main text.
Since $\Delta_z'$ is real, Eq.~\eqref{eq:phase_locking} requires
$\left(
\mathcal C_r-\mathcal C_i
\right)
\sin(2\phi)
=
0$.
Equivalently,
\begin{equation}
\left(
\frac{g_a\lambda_a}{\omega_a}
+
\frac{g_b\lambda_b}{\omega_b}
\right)
\sin(2\phi)
=
0.
\label{eq:phase_condition_bare}
\end{equation}
Equation~\eqref{eq:phase_locking} provides a unified description of the
superradiant ordering. 
For $\mathcal C_r\neq\mathcal C_i$, the condensate phase is therefore
restricted to $\phi=0,\ \pi,\ \pm\frac{\pi}{2}.$
The solutions $\phi=0,\pi$ correspond to purely real condensates
\begin{equation}
	\alpha_r
	=
	\pm\frac{\sin\theta_r}{2\omega_a}A_+,
	\qquad
	\beta_r
	=
	\pm\frac{\sin\theta_r}{2\omega_b}B_+,
	\label{eq:SRr_order_parameters}
\end{equation}
and define the $\mathrm{SR}_r$ branch. Equation~\eqref{eq:phase_locking}
then gives $\Delta_r=\mathcal C_r.$
By contrast, $\phi=\pm\pi/2$ gives purely imaginary condensates 
\begin{equation}
	\alpha_i
	=
	\pm i\frac{\sin\theta_i}{2\omega_a}A_-,
	\qquad
	\beta_i
	=
	\pm i\frac{\sin\theta_i}{2\omega_b}B_-,
	\label{eq:SRi_order_parameters}
\end{equation}
and
defines the $\mathrm{SR}_i$ branch, for which $\Delta_i=\mathcal C_i.$
Thus, the two phase-locked superradiant branches are directly associated
with the coordinate-like and momentum-like collective coupling channels
$\mathcal C_r$ and $\mathcal C_i$, respectively.

At the normal-to-superradiant boundary, the condensate amplitudes vanish,
so that $\Delta_r,\Delta_i\rightarrow\Delta_z$. Equations
~\eqref{eq:phase_locking} therefore reduce to $\mathcal C_r=\Delta_z$, $\mathcal C_i=\Delta_z$,
which reproduce the mean-field and BdG soft-mode conditions for the
$N$--$\mathrm{SR}_r$ and $N$--$\mathrm{SR}_i$ transitions,
respectively. A distinct situation occurs when $\mathcal C_r=\mathcal C_i,$
or equivalently $\frac{g_a\lambda_a}{\omega_a}
+
\frac{g_b\lambda_b}{\omega_b}
=
0.$
In this case, the phase-dependent term in
Eq.~\eqref{eq:phase_locking} vanishes, and the
$\mathrm{SR}_r$ and $\mathrm{SR}_i$ branches become degenerate.
For finite counter-rotating-wave couplings, this degeneracy occurs
without restoration of a continuous symmetry and defines the
first-order boundary between the two superradiant phases. This finite-anisotropy degeneracy should be distinguished from the
$U(1)$-symmetric limit $\lambda_a=\lambda_b=0$. In the latter case,
$\mathcal C_r=\mathcal C_i$ holds identically as a consequence of
continuous-symmetry restoration, so that the condensate phase remains
unlocked and the superradiant states form a continuously degenerate
manifold with an associated Goldstone-like mode. Hence, the three conditions
\begin{equation}
\mathcal C_r=\Delta_z,
\qquad
\mathcal C_i=\Delta_z,
\qquad
\mathcal C_r=\mathcal C_i,
\end{equation}
determine the two continuous normal-to-superradiant boundaries and the
first-order boundary between the two superradiant branches, respectively.

\section{Low-Energy Effective Hamiltonians in the Superradiant Phases}
	\label{app:lowenergy-superradiantphase}

The phase-locking analysis above determines the self-consistent displaced
backgrounds of the two superradiant branches, $\mathrm{SR}_r$ and
$\mathrm{SR}_i$, characterized by the condensate amplitudes
$(\alpha_\mu,\beta_\mu)$, with
$\mu=r,i$.
The $\mathrm{SR}_r$ branch is associated with a real condensate and
coordinate-like ordering, whereas the $\mathrm{SR}_i$ branch is associated
with a purely imaginary condensate and momentum-like ordering.
For each superradiant branch, we expand the bosonic operators around the
corresponding self-consistent displaced background as $\hat a=\alpha_\mu+\hat c$ and $\hat b=\beta_\mu+\hat d$,
where $\hat c$ and $\hat d$ describe quantum fluctuations about the
phase-locked condensate.

For the subsequent SW analysis, it is convenient to express
these fluctuations in the canonical quadrature representation,
\begin{equation}
x_c=
\frac{\hat c+\hat c^\dagger}{\sqrt{2}},
\qquad
p_c=
\frac{i(\hat c^\dagger-\hat c)}{\sqrt{2}},
\label{eq:cd_quadrature_a}
\end{equation}
and
\begin{equation}
x_d=
\frac{\hat d+\hat d^\dagger}{\sqrt{2}},
\qquad
p_d=
\frac{i(\hat d^\dagger-\hat d)}{\sqrt{2}}.
\label{eq:cd_quadrature_b}
\end{equation}
Equivalently, the original bosonic quadratures are shifted according to
\begin{align}
x_a
&=
x_c+\sqrt{2}\,\mathrm{Re}\,\alpha_\mu,
&
p_a
&=
p_c+\sqrt{2}\,\mathrm{Im}\,\alpha_\mu,
\nonumber\\
x_b
&=
x_d+\sqrt{2}\,\mathrm{Re}\,\beta_\mu,
&
p_b
&=
p_d+\sqrt{2}\,\mathrm{Im}\,\beta_\mu.
\label{eq:SR_quadrature_shift}
\end{align}
Thus, the phase-locking analysis fixes the displaced background
$(\alpha_\mu,\beta_\mu)$, while the quadratures
$(x_c,p_c,x_d,p_d)$ describe the collective fluctuations around that
background. This representation is particularly convenient for separating
the coordinate-like and momentum-like fluctuation sectors in the
low-energy theory. 
In the large-frequency-ratio regime, the dressed-spin excitation remains
energetically separated from these low-energy bosonic fluctuations.
The high-energy dressed-spin sector can therefore be perturbatively
eliminated through a SW transformation, yielding the
quadratic effective Hamiltonian for each superradiant branch.

\subsection{Real-Displacement Superradiant Branch}

We first consider the $\mathrm{SR}_r$ branch, for which the
self-consistent condensate amplitudes are purely real, $\alpha_r,\beta_r\in\mathbb{R}$.
Accordingly, the displacement occurs entirely in the coordinate sector,
$x_a=x_c+\sqrt{2}\alpha_r$, $x_b=x_d+\sqrt{2}\beta_r$,
while $p_a=p_c$, $p_b=p_d$.

The condensate generates an effective transverse field along the
$\sigma_x$ direction, so that the displacement-dependent spin sector is
\begin{equation}
\hat H_{\rm spin}^{r}
=
\frac{\Delta_z}{2}\hat\sigma_z
+
\Gamma_r\hat\sigma_x,
\label{eq:Hspin_r}
\end{equation}
where $\Gamma_r
=
A_+\alpha_r+B_+\beta_r$.
The corresponding dressed-spin gap is $\Delta_r
=
\sqrt{
\Delta_z^2+4\Gamma_r^2
}.$

Rotating the spin quantization axis along the effective field gives
\begin{equation}
\hat H_{\rm spin}^{r}
=
\frac{\Delta_r}{2}\hat\tau_z,
\end{equation}
with $\cos\theta_r
=
\frac{\Delta_z}{\Delta_r}$, $\sin\theta_r
=
\frac{2\Gamma_r}{\Delta_r}.$
The original spin operators are related to the dressed-spin operators by
\begin{equation}
\hat\sigma_x
=
\sin\theta_r\,\hat\tau_z
+
\cos\theta_r\,\hat\tau_x,
\qquad
\hat\sigma_z
=
\cos\theta_r\,\hat\tau_z
-
\sin\theta_r\,\hat\tau_x,
\qquad
\hat\sigma_y
=
\hat\tau_y.
\label{eq:spin_rotation_r}
\end{equation}

Substituting the displacement and spin rotation into the Hamiltonian,
the terms linear in the coordinate fluctuations and diagonal in the
dressed-spin basis are
\begin{align}
\hat H_{\rm lin}^{\mathrm{SR}_r}
=&\,
\left(
\sqrt{2}\omega_a\alpha_r
+
\frac{A_+}{\sqrt{2}}
\sin\theta_r\,\hat\tau_z
\right)x_c
\nonumber\\
&+
\left(
\sqrt{2}\omega_b\beta_r
+
\frac{B_+}{\sqrt{2}}
\sin\theta_r\,\hat\tau_z
\right)x_d .
\label{eq:linear_SRr}
\end{align}
Projecting onto the dressed-spin ground state,
$\langle\hat\tau_z\rangle=-1$, and requiring the resulting linear
terms to vanish gives the self-consistent condensate amplitudes
\begin{equation}
\alpha_r
=
\frac{A_+}{2\omega_a}\sin\theta_r,
\qquad
\beta_r
=
\frac{B_+}{2\omega_b}\sin\theta_r.
\label{eq:alpha_beta_r}
\end{equation}
These relations coincide with the real-displacement solution obtained
from the phase-locking analysis above.

After imposing the self-consistency conditions, the linear terms
associated with variations of the condensate amplitudes vanish.
The remaining off-diagonal spin--boson couplings describe quantum
fluctuations around the self-consistent displaced background.
For the $\mathrm{SR}_r$ branch, the displaced Hamiltonian becomes
\begin{align}
\hat H_{\rm disp}^{\mathrm{SR}_r}
=&\,
\bar E_{\mathrm{SR}_r}
+
\omega_a \hat c^\dagger \hat c
+
\omega_b \hat d^\dagger \hat d
+
\frac{\Delta_r}{2}
\left(
\hat\tau_z+1
\right)
\nonumber\\
&+
\frac{A_+}{\sqrt{2}}
\cos\theta_r\,\hat\tau_x x_c
+
\frac{B_+}{\sqrt{2}}
\cos\theta_r\,\hat\tau_x x_d
\nonumber\\
&-
\frac{A_-}{\sqrt{2}}
\hat\tau_y p_c
-
\frac{B_-}{\sqrt{2}}
\hat\tau_y p_d ,
\label{eq:Hdisp_SRr}
\end{align}
where
\begin{equation}
\bar E_{\mathrm{SR}_r}
=
\omega_a\alpha_r^2
+
\omega_b\beta_r^2
-
\frac{\Delta_r}{2}
\label{eq:Ebar_SRr}
\end{equation}
is the mean-field energy of the self-consistent
$\mathrm{SR}_r$ background.
Equation~\eqref{eq:Hdisp_SRr} shows that the fluctuations around the
$\mathrm{SR}_r$ condensate retain two distinct coupling channels.
The coordinate-like coupling is renormalized by
$\cos\theta_r$, whereas the momentum-like coupling remains unrenormalized.
In the large-frequency-ratio regime, the high-energy dressed-spin
excitation can therefore be perturbatively eliminated through a
SW transformation, yielding the quadratic low-energy
Hamiltonian for the collective fluctuations in the $\mathrm{SR}_r$ phase.

\subsection{Imaginary-Displacement Superradiant Branch}

We next consider the $\mathrm{SR}_i$ branch, for which the
self-consistent condensate amplitudes are purely imaginary,
\begin{equation}
\alpha_{\mathrm{SR}_i}=i\alpha_i,
\qquad
\beta_{\mathrm{SR}_i}=i\beta_i,
\qquad
\alpha_i,\beta_i\in\mathbb{R}.
\label{eq:SRi_displacement}
\end{equation}
Accordingly, the displacement occurs entirely in the momentum sector,
\begin{equation}
p_a=p_c+\sqrt{2}\alpha_i,
\qquad
p_b=p_d+\sqrt{2}\beta_i,
\end{equation}
while $x_a=x_c$, $x_b=x_d$.

The condensate generates an effective transverse field along the
$\sigma_y$ direction, so that the displacement-dependent spin sector
takes the form
\begin{equation}
\hat H_{\rm spin}^{i}
=
\frac{\Delta_z}{2}\hat\sigma_z
-
\Gamma_i\hat\sigma_y,
\label{eq:Hspin_i}
\end{equation}
where $\Gamma_i=
A_-\alpha_i+B_-\beta_i.$
The corresponding dressed-spin gap is
\begin{equation}
\Delta_i
=
\sqrt{
\Delta_z^2+4\Gamma_i^2
}.
\label{eq:Delta_i_def}
\end{equation}

Rotating the spin quantization axis along the effective field gives
\begin{equation}
\hat H_{\rm spin}^{i}
=
\frac{\Delta_i}{2}\hat\eta_z,
\end{equation}
with $\cos\theta_i
=
\frac{\Delta_z}{\Delta_i},$ $\sin\theta_i
=
\frac{2\Gamma_i}{\Delta_i}.$
The original spin operators are related to the dressed-spin operators by
\begin{equation}
\hat\sigma_y
=
-\sin\theta_i\,\hat\eta_z
+
\cos\theta_i\,\hat\eta_y,
\qquad
\hat\sigma_z
=
\cos\theta_i\,\hat\eta_z
+
\sin\theta_i\,\hat\eta_y,
\qquad
\hat\sigma_x
=
\hat\eta_x.
\label{eq:spin_rotation_i}
\end{equation}

Substituting the displacement and spin rotation into the Hamiltonian,
the terms linear in the momentum fluctuations and diagonal in the
dressed-spin basis are
\begin{align}
\hat H_{\rm lin}^{\mathrm{SR}_i}
=&\,
\left(
\sqrt{2}\omega_a\alpha_i
+
\frac{A_-}{\sqrt{2}}
\sin\theta_i\,\hat\eta_z
\right)p_c
\nonumber\\
&+
\left(
\sqrt{2}\omega_b\beta_i
+
\frac{B_-}{\sqrt{2}}
\sin\theta_i\,\hat\eta_z
\right)p_d .
\label{eq:linear_SRi}
\end{align}
Projecting onto the dressed-spin ground state,
$\langle\hat\eta_z\rangle=-1$, and requiring the resulting linear
terms to vanish gives the self-consistent condensate amplitudes
\begin{equation}
\alpha_i
=
\frac{A_-}{2\omega_a}\sin\theta_i,
\qquad
\beta_i
=
\frac{B_-}{2\omega_b}\sin\theta_i.
\label{eq:alpha_beta_i}
\end{equation}
These relations coincide with the imaginary-displacement solution
obtained from the phase-locking analysis above.

After imposing the self-consistency conditions, the linear terms
associated with variations of the condensate amplitudes vanish.
The remaining off-diagonal spin--boson couplings describe quantum
fluctuations around the self-consistent displaced background.
The Hamiltonian describing fluctuations
around the $\mathrm{SR}_i$ background can then be written as
\begin{align}
\hat H_{\rm disp}^{\mathrm{SR}_i}
=&\,
\bar E_{\mathrm{SR}_i}
+
\omega_a\hat c^\dagger\hat c
+
\omega_b\hat d^\dagger\hat d
+
\frac{\Delta_i}{2}
\left(
\hat\eta_z+1
\right)
\nonumber\\
&-
\frac{A_+}{\sqrt{2}}
\hat\eta_x x_c
-
\frac{B_+}{\sqrt{2}}
\hat\eta_x x_d
\nonumber\\
&-
\frac{A_-}{\sqrt{2}}
\cos\theta_i\,\hat\eta_y p_c
-
\frac{B_-}{\sqrt{2}}
\cos\theta_i\,\hat\eta_y p_d ,
\label{eq:Hdisp_SRi}
\end{align}
where
\begin{equation}
\bar E_{\mathrm{SR}_i}
=
\omega_a\alpha_i^2
+
\omega_b\beta_i^2
-
\frac{\Delta_i}{2}
\label{eq:Ebar_SRi}
\end{equation}
is the mean-field energy of the self-consistent
$\mathrm{SR}_i$ background.

Equation~\eqref{eq:Hdisp_SRi} shows that the fluctuations around the
$\mathrm{SR}_i$ condensate retain two distinct coupling channels.
In contrast to the $\mathrm{SR}_r$ branch, the momentum-like coupling
is renormalized by $\cos\theta_i$, whereas the coordinate-like coupling
remains unrenormalized. In the large-frequency-ratio regime, the
high-energy dressed-spin excitation can therefore be perturbatively
eliminated through a SW transformation, yielding the
quadratic low-energy Hamiltonian for the collective fluctuations in the
$\mathrm{SR}_i$ phase.

\subsection{Schrieffer--Wolff Effective Low-Energy Hamiltonians}

Having established the self-consistent displaced backgrounds and the
corresponding dressed-spin bases, we now eliminate the high-energy
dressed-spin excitations through a SW transformation.
For both superradiant branches, the remaining off-diagonal spin--boson
coupling can be written in the unified form
\begin{equation}
\hat V_\mu
=
\hat s_x^{(\mu)}X_\mu
+
\hat s_y^{(\mu)}P_\mu,
\qquad
\mu=\mathrm{SR}_r,\mathrm{SR}_i,
\label{eq:Vmu}
\end{equation}
where
\begin{equation}
\hat s_\alpha^{(\mathrm{SR}_r)}
=
\hat\tau_\alpha,
\qquad
\hat s_\alpha^{(\mathrm{SR}_i)}
=
\hat\eta_\alpha.
\label{eq:dressed_spin_mu}
\end{equation}
The corresponding dressed-spin Hamiltonian is
\begin{equation}
\hat H_{0,\mu}
=
\frac{\Delta_\mu}{2}\hat s_z^{(\mu)}.
\label{eq:H0mu}
\end{equation}

For the $\mathrm{SR}_r$ branch, the fluctuation operators are
\begin{equation}
X_r
=
\frac{\cos\theta_r}{\sqrt{2}}
\left(
A_+x_c+B_+x_d
\right),
\qquad
P_r
=
-\frac{1}{\sqrt{2}}
\left(
A_-p_c+B_-p_d
\right),
\label{eq:XP_r}
\end{equation}
whereas for the $\mathrm{SR}_i$ branch they are
\begin{equation}
X_i
=
-\frac{1}{\sqrt{2}}
\left(
A_+x_c+B_+x_d
\right),
\qquad
P_i
=
-\frac{\cos\theta_i}{\sqrt{2}}
\left(
A_-p_c+B_-p_d
\right).
\label{eq:XP_i}
\end{equation}

The two branches can therefore be treated by the same unitary SW
transformation,
\begin{equation}
\hat U_\mu
=
e^{\hat S_\mu},
\qquad
\hat S_\mu
=
\frac{i}{\Delta_\mu}
\left[
\hat s_y^{(\mu)}X_\mu
-
\hat s_x^{(\mu)}P_\mu
\right],
\label{eq:SWUnified}
\end{equation}
for which
\begin{equation}
[\hat S_\mu,\hat H_{0,\mu}]
=
-\hat V_\mu.
\label{eq:SWConditionUnified}
\end{equation}
Expanding the transformed Hamiltonian to second order in the
light--matter couplings and projecting onto the low-energy dressed-spin
sector yields quadratic effective Hamiltonians for the collective
bosonic fluctuations.

For the $\mathrm{SR}_r$ branch,
$\langle\hat\tau_z\rangle=-1$, and the resulting effective Hamiltonian is
\begin{align}
\hat H_{\rm eff}^{\mathrm{SR}_r}
={}&
\bar E_{\mathrm{SR}_r}
+
\frac{\omega_a}{2}
\left(
x_c^2+p_c^2-1
\right)
+
\frac{\omega_b}{2}
\left(
x_d^2+p_d^2-1
\right)
\nonumber\\
&-
\frac{\cos^2\theta_r}{2\Delta_r}
\left(
A_+x_c+B_+x_d
\right)^2
\nonumber\\
&-
\frac{1}{2\Delta_r}
\left(
A_-p_c+B_-p_d
\right)^2 .
\label{eq:HeffSRr}
\end{align}
Here, $\bar E_{\mathrm{SR}_r}$ denotes the mean-field energy of the
self-consistent $\mathrm{SR}_r$ background defined above. The
coordinate-like fluctuation channel is renormalized by
$\cos^2\theta_r$, whereas the momentum-like channel retains its full
coupling strength.

For the $\mathrm{SR}_i$ branch,
$\langle\hat\eta_z\rangle=-1$, yielding
\begin{align}
\hat H_{\rm eff}^{\mathrm{SR}_i}
={}&
\bar E_{\mathrm{SR}_i}
+
\frac{\omega_a}{2}
\left(
x_c^2+p_c^2-1
\right)
+
\frac{\omega_b}{2}
\left(
x_d^2+p_d^2-1
\right)
\nonumber\\
&-
\frac{1}{2\Delta_i}
\left(
A_+x_c+B_+x_d
\right)^2
\nonumber\\
&-
\frac{\cos^2\theta_i}{2\Delta_i}
\left(
A_-p_c+B_-p_d
\right)^2 .
\label{eq:HeffSRi}
\end{align}
Here, $\bar E_{\mathrm{SR}_i}$ denotes the mean-field energy of the
self-consistent $\mathrm{SR}_i$ background defined above. In contrast
to the $\mathrm{SR}_r$ branch, the momentum-like fluctuation channel is
renormalized by $\cos^2\theta_i$, whereas the coordinate-like channel
retains its full coupling strength.

\section{Unified BdG Formulation and Ground-State Energies}
\label{app:UnifiedBdG}

The quadratic effective Hamiltonians derived above can be written in the
unified form as
\begin{align}
\hat H_{\rm eff}^{\mu}
={}&
\bar E_{\mu}
+
\frac{\omega_a}{2}
\left(
x_c^2+p_c^2-1
\right)
+
\frac{\omega_b}{2}
\left(
x_d^2+p_d^2-1
\right)
\nonumber\\
&-
\frac{C_x^\mu}{2}
\left(
A_+x_c+B_+x_d
\right)^2
-
\frac{C_p^\mu}{2}
\left(
A_-p_c+B_-p_d
\right)^2,
\label{eq:HeffUnifiedQuadrature}
\end{align}
with the phase
dependence encoded in the coefficients $C_x^\mu$ and $C_p^\mu$. Explicitly,
\begin{equation}
C_x^N=C_p^N=\frac{1}{\Delta_z},
\label{eq:Cxp_N}
\end{equation}
\begin{equation}
C_x^{\rm SR-r}
=
\frac{\cos^2\theta_r}{\Delta_r},
\qquad
C_p^{\rm SR-r}
=
\frac{1}{\Delta_r},
\label{eq:Cxp_SRr}
\end{equation}
and
\begin{equation}
C_x^{\rm SR-i}
=
\frac{1}{\Delta_i},
\qquad
C_p^{\rm SR-i}
=
\frac{\cos^2\theta_i}{\Delta_i}.
\label{eq:Cxp_SRi}
\end{equation}
Thus, the normal phase treats the coordinate-like and momentum-like
fluctuation sectors on equal footing, whereas the two superradiant
branches exhibit complementary quadrature-selective renormalizations. Equations~\eqref{eq:HeffUnifiedQuadrature} show that the phase
dependence is fully encoded in $C_x^\mu$ and $C_p^\mu$, while the
quadratic structure of the effective Hamiltonian remains unchanged.
We can therefore apply the same BdG diagonalization introduced for the
normal phase to superradiant branches, yielding the collective excitation
frequencies $\Omega_{\mu,\pm}$.
The corresponding Bogoliubov-corrected ground-state energy is
\begin{equation}
E_{\rm GS}^{\mu}
=
\bar E_{\mu}
+
\frac{1}{2}
\left(
\Omega_{\mu,+}
+
\Omega_{\mu,-}
-
\omega_a
-
\omega_b
\right),
\qquad
\mu=N,\mathrm{SR}_r,\mathrm{SR}_i,
\label{eq:EGSfull}
\end{equation}
with the two positive Bogoliubov branches
\begin{equation}
\Omega_{\mu,\pm}
=
\sqrt{
\frac{1}{2}
\left[
T_\mu
\pm
\sqrt{
T_\mu^2-4\mathcal D_\mu
}
\right]
}.
\label{eq:BdGfreq}
\end{equation}
The lower branch $\Omega_{\mu,-}$ determines the local stability of the
corresponding solution, with $\Omega_{\mu,-}=0$ marking its soft-mode
instability.

The corresponding coordinate- and momentum-sector stiffness matrices are
\begin{equation}
K_x^\mu
=
\begin{pmatrix}
\displaystyle
\omega_a-C_x^\mu A_+^2
&
\displaystyle
-C_x^\mu A_+B_+
\\[2mm]
\displaystyle
-C_x^\mu A_+B_+
&
\displaystyle
\omega_b-C_x^\mu B_+^2
\end{pmatrix},
\label{eq:KxUnified}
\end{equation}
and
\begin{equation}
K_p^\mu
=
\begin{pmatrix}
\displaystyle
\omega_a-C_p^\mu A_-^2
&
\displaystyle
-C_p^\mu A_-B_-
\\[2mm]
\displaystyle
-C_p^\mu A_-B_-
&
\displaystyle
\omega_b-C_p^\mu B_-^2
\end{pmatrix}.
\label{eq:KpUnified}
\end{equation}

The squared collective excitation frequencies are the eigenvalues of
$K_p^\mu K_x^\mu$. Defining
\begin{equation}
T_\mu
=
{\rm Tr}\!\left(
K_p^\mu K_x^\mu
\right),
\qquad
\mathcal D_\mu
=
\det\!\left(
K_p^\mu K_x^\mu
\right),
\label{eq:TDmu}
\end{equation}

The physical ground-state energy is then obtained by minimizing over the
three locally stable branches,
\begin{equation}
E_{\rm GS}
=
\min
\left\{
E_{\rm GS}^{N},
E_{\rm GS}^{\mathrm{SR}_r},
E_{\rm GS}^{\mathrm{SR}_i}
\right\}.
\label{eq:GlobalGS}
\end{equation}
Thus, the same quadratic formulation determines both the collective
excitation spectra and the quantum-corrected ground-state energies of
the normal and superradiant phases.

\section{Soft-Mode Conditions and Instability Channels}
\label{app:SoftModeConditions}

Within the unified BdG formulation, the local stability of each branch,
$\mu=N,\mathrm{SR}_r,\mathrm{SR}_i$, is governed by the lower collective
excitation frequency $\Omega_{\mu,-}$. The soft-mode condition
\begin{equation}
\Omega_{\mu,-}=0
\label{eq:SoftModeCondition}
\end{equation}
is equivalent to
\begin{equation}
\mathcal D_\mu
=
\det\!\left(K_p^\mu K_x^\mu\right)
=
\det K_p^\mu\,\det K_x^\mu
=
0.
\label{eq:SoftDetCondition}
\end{equation}
Thus, the instability can arise from either the coordinate-like sector,
$\det K_x^\mu=0$, or the momentum-like sector,
$\det K_p^\mu=0$.

Using the definitions of $\mathcal C_r$ and $\mathcal C_i$ introduced
above, the determinants of the stiffness matrices take the compact form
\begin{equation}
\det K_x^\mu
=
\omega_a\omega_b
\left(
1-C_x^\mu\mathcal C_r
\right),
\qquad
\det K_p^\mu
=
\omega_a\omega_b
\left(
1-C_p^\mu\mathcal C_i
\right).
\label{eq:detKxp}
\end{equation}
Consequently,
\begin{equation}
\mathcal D_\mu
=
\omega_a^2\omega_b^2
\left(
1-C_x^\mu\mathcal C_r
\right)
\left(
1-C_p^\mu\mathcal C_i
\right),
\label{eq:DmuUnified}
\end{equation}
and the unified soft-mode condition becomes
\begin{equation}
\left(
1-C_x^\mu\mathcal C_r
\right)
\left(
1-C_p^\mu\mathcal C_i
\right)
=
0,
\qquad
\mu=N,\mathrm{SR}_r,\mathrm{SR}_i.
\label{eq:UnifiedSoftCondition}
\end{equation}

For the normal phase,
$C_x^N=C_p^N=1/\Delta_z$, so that
\begin{equation}
\mathcal D_N
=
\omega_a^2\omega_b^2
\left(
1-\frac{\mathcal C_r}{\Delta_z}
\right)
\left(
1-\frac{\mathcal C_i}{\Delta_z}
\right),
\label{eq:DN}
\end{equation}
and the two soft-mode conditions are
\begin{equation}
\mathcal C_r=\Delta_z,
\qquad
\mathcal C_i=\Delta_z.
\label{eq:NormalSoftConditions}
\end{equation}
These coincide with the mean-field critical conditions for the
$N$--$\mathrm{SR}_r$ and $N$--$\mathrm{SR}_i$ transitions,
respectively.

For the $\mathrm{SR}_r$ branch,
\begin{equation}
\mathcal D_{\mathrm{SR}_r}
=
\omega_a^2\omega_b^2
\left(
1-
\frac{\cos^2\theta_r}{\Delta_r}
\mathcal C_r
\right)
\left(
1-
\frac{\mathcal C_i}{\Delta_r}
\right),
\label{eq:Dr}
\end{equation}
which gives
\begin{equation}
\mathcal C_r
=
\frac{\Delta_r}{\cos^2\theta_r},
\qquad
\mathcal C_i
=
\Delta_r.
\label{eq:SRrSoftConditions}
\end{equation}

Similarly, for the $\mathrm{SR}_i$ branch,
\begin{equation}
\mathcal D_{\mathrm{SR}_i}
=
\omega_a^2\omega_b^2
\left(
1-
\frac{\mathcal C_r}{\Delta_i}
\right)
\left(
1-
\frac{\cos^2\theta_i}{\Delta_i}
\mathcal C_i
\right),
\label{eq:Di}
\end{equation}
yielding
\begin{equation}
\mathcal C_r
=
\Delta_i,
\qquad
\mathcal C_i
=
\frac{\Delta_i}{\cos^2\theta_i}.
\label{eq:SRiSoftConditions}
\end{equation}

Equations~\eqref{eq:NormalSoftConditions},
\eqref{eq:SRrSoftConditions}, and
\eqref{eq:SRiSoftConditions} give the soft-mode conditions for the
normal and two superradiant branches within the same BdG framework.
The coordinate-like and momentum-like instability channels are therefore
encoded by $\det K_x^\mu$ and $\det K_p^\mu$, respectively, with their
phase dependence determined entirely by $C_x^\mu$ and $C_p^\mu$.

\end{document}